\def\year{2020}\relax

\documentclass[sigconf]{acmart}

\usepackage{multicol}
\usepackage[utf8]{inputenc}
\usepackage{blindtext}
\usepackage[T1]{fontenc}
\usepackage{natbib}
\usepackage{graphics}
\usepackage{hyperref}
\usepackage{dcolumn}
\newcolumntype{d}[1]{D{.}{.}{-1}}
\newcolumntype{n}[1]{D{,}{,}{-1}}
\usepackage{pdflscape}

\newcommand{\xhdr}[1]{\paragraph*{\bf #1}}
\title{The Structure of U.S. College Networks on Facebook}

\author{Jan Overgoor}
  \affiliation{\institution{Facebook}}
  \email{janovergoor@gmail.com}
\author{Bogdan State}
  \affiliation{\institution{Facebook}}
  \email{bogdanstate@fb.com}
\author{Lada Adamic}
  \affiliation{\institution{Facebook}}
  \email{ladamic@fb.com}

\begin{document}

\begin{abstract}
Anecdotally, social connections made in university have life-long impact.
Yet knowledge of social networks formed in college remains episodic,
  due in large part to the difficulty and expense involved in collecting a suitable dataset for comprehensive analysis.
To advance and systematize insight into college social networks,
  we describe a dataset of the largest online social network platform used by college students in the United States.
We combine de-identified and aggregated Facebook data with College Scorecard data,
  campus-level information provided by U.S. Department of Education,
  to produce a dataset covering the 2008-2015 entry year cohorts for 1,159 U.S. colleges and universities, spanning 7.6 million students. 
To perform the difficult task of comparing these networks of different sizes we develop a new methodology.
We compute features over sampled ego-graphs,
  train binary classifiers for every pair of graphs,
  and operationalize distance between graphs as predictive accuracy.
Social networks of different year cohorts at the same school
  are structurally more similar to one another than to cohorts at other schools.
Networks from similar schools have similar structures,
  with the public/private and graduation rate dimensions being the most distinguishable.
We also relate school types to specific outcomes.
For example, students at private schools have larger networks that are more clustered and with higher homophily by year.
Our findings may help illuminate the role that colleges play in shaping social networks which partly persist throughout people's lives.
\end{abstract}

\maketitle

\section{Introduction}

In the United States, more than half of the adult population
   has attended an institution of  higher education \citep{ryan16}.
The time spent there is commonly perceived to be formative for life.
For one, colleges are thought of as engines of social mobility \citep{chetty17}.
For many, college is the first exposure to a larger and
  more diverse social environment than where they grew up,
  which supports the finding of work \citep{granovetter73}.
It is common for people to find their partners or
  close friends during college \citep{arum08},
  which contributes to social stratification in U.S. society more generally \citep{gerber08}.

Although college students are perhaps the most intensely studied population by social scientists,
  it has proven hard to base claims on comprehensive data of actual social networks at scale \citep{biancani13}.
Probably the most important reason for this is the complexity of the data involved.
Investigating the social networks even within a single class is a daunting task \citep{mcfarland01}.
There are many relevant actors and social contexts to consider,
  and the data is expensive to collect, error prone, and often changes rapidly.
These difficulties have led to most studies being
  focused on limited settings, and often using at most few static snapshots of data.
Data from online social networks helps address some of these limitations,
  especially around scale and temporal dynamics.
Because Facebook has historically been popular with college students
  (the platform having originally only been available on US college campuses),
  it provides a natural context to study the social networks of college students
  for this period of time.
However, most prior work based on Facebook data has been limited to a small number of mostly elite institutions and/or relied on a single static snapshot of the social connections.

In this work, we present a population-level descriptive overview of the structure of online social networks formed in college,
  based on de-identified and aggregated data from Facebook.
Within this scope we study 7,660 distinct class cohorts associated with 1,159 U.S. institutions of higher education who began their studies during the period 2008-2015.
Covered are a total of 7.6 million students, with 552.6 million edges between them.
We connect the structural features of the resulting school networks
  with data on school characteristics as provided by the U.S. Department of Education.
This dataset allows us to characterize how the online social networks of college students vary by the school they attend,
  and to do so for a wide range of institution types.

We do so in two ways.
First, we compute similarities between school graphs, and relate them to differences in school types.
Comparing graphs of different sizes is a hard problem \citep{shalizi13},
  so we employ a new methodology for doing so.
We measure the distance between two graphs of different sizes
  as the predictive accuracy of a binary classifier
  trained on features over sampled ego-graphs.
The structure of the resulting pairwise distance matrix is correlated with school characteristics,
  such as whether the institution is public or private, the graduation rate, and Greek life participation.
This indicates that students at different kinds of schools form consistently different social networks.

Next, we look at how structural features of the social networks, like average degree, clustering and homophily, relate to characteristics of schools.
For example, we find that, accounting for size, graduation rate, and admission rate,
  students at private schools have a larger network than those at similar public schools.
Their networks are also more clustered and more segregated by year.
Similarly, students at schools with high Greek participation have larger networks,
  with more clustering, more mixing across years, and more gender homophily.
Students at Historically Black Colleges and Universities make significantly more
  connections within their college, while those at women's-only institutions have fewer, perhaps because cross-gender friendships occur outside of the college.
We make the summary statistics of the class graphs available to be downloaded for future study.

Our work provides the first population-level overview of the online social networks of U.S. college students.
We aim to help fill the gap between the anecdotally important role that college plays in people's lives, and comprehensive data to help study this important social setting.
While our findings are observational,
they provide a first opportunity to 
study a population of college networks, thus allowing us to tease apart different
aspects of network ecology. 
Since heterogeneity in colleges gives rise to heterogeneity in network structure, and network
structure has been tied to social and economic outcomes, understanding the
structure is a step toward understanding how the college environment
helps shape an important part of people's lives, their social networks.

\subsection{Related Work}

Social networks in schools and their relation to outcomes
  have long preoccupied education researchers and sociologists \citep{hanifan16,french48,moreno34}.
The role of the mechanisms of propinquity and homophily in network formation are often recurring themes.
Propinquity is the tendency for social networks to be spatially organized,
  with proximity a key factor influencing the likelihood of social tie formation.
This is relevant with respect to shared foci, like dormitories \citep{festinger50}, classes \citep{kossinets06},
  and extracurricular activities \citep{van03}.
Homophily is the tendency to associate with others who are similar \citep{mcpherson01,currarini09},
  and in educational settings can occur along dimensions like race, gender, and socio-economic status \citep{marmaros06,godley08,wimmer10}.
However, as pointed out by Biancani and McFarland (\citeyear{biancani13}), the literature still lacks a comprehensive descriptive account of university social networks.
In contrast, the comparative study of social networks in high schools has advanced more
  thanks to the longitudinal ``Add Health'' dataset \citep{harris08},
  which collected data from over 90,000 individuals who were enrolled in middle school or high school in the US during the 1994-95 academic year.
This includes research on homophily \citep{joyner00},
  the propinquity of extra-curricular activities \citep{schaefer11}, 
  and the relation of various behaviors to network structure.
One particular work takes a similar approach to ours,
  in that they relate structural elements of school networks to school covariates \cite{mcfarland14}.

Because of the natural connection between Facebook and higher education,
  it has been used as a data source for multiple small- and medium-scale studies of college social networks.
This work has highlighted the strong role of race homophily \citep{mayer08,wimmer10},
  as well as homophily by high school \citep{traud11}
  and student residences \citep{traud12}.
By measuring Facebook ``likes'' as a proxy for taste,
  researchers have identified taste-similarity among friends \citep{lewis08},
  and that this is more likely due to homophily rather than social influence \citep{lewis12}.
However, due to limited data availability,
  these works have generally been focused on either a single institution 
  or relied on a single static snapshot of the social graphs. 
This has prevented insight into patterns over time,
  as well as variation across schools.

\section{Data construction}

In this section, we describe the construction of the college networks dataset.
We combined data from two sources: self-reported demographic information and information from the Facebook platform,
and external information related to institutions of higher education in the United States.
All data were de-identified and analyzed in aggregate.

\begin{figure}[t]
  \centering
  \includegraphics[width=0.9\columnwidth]{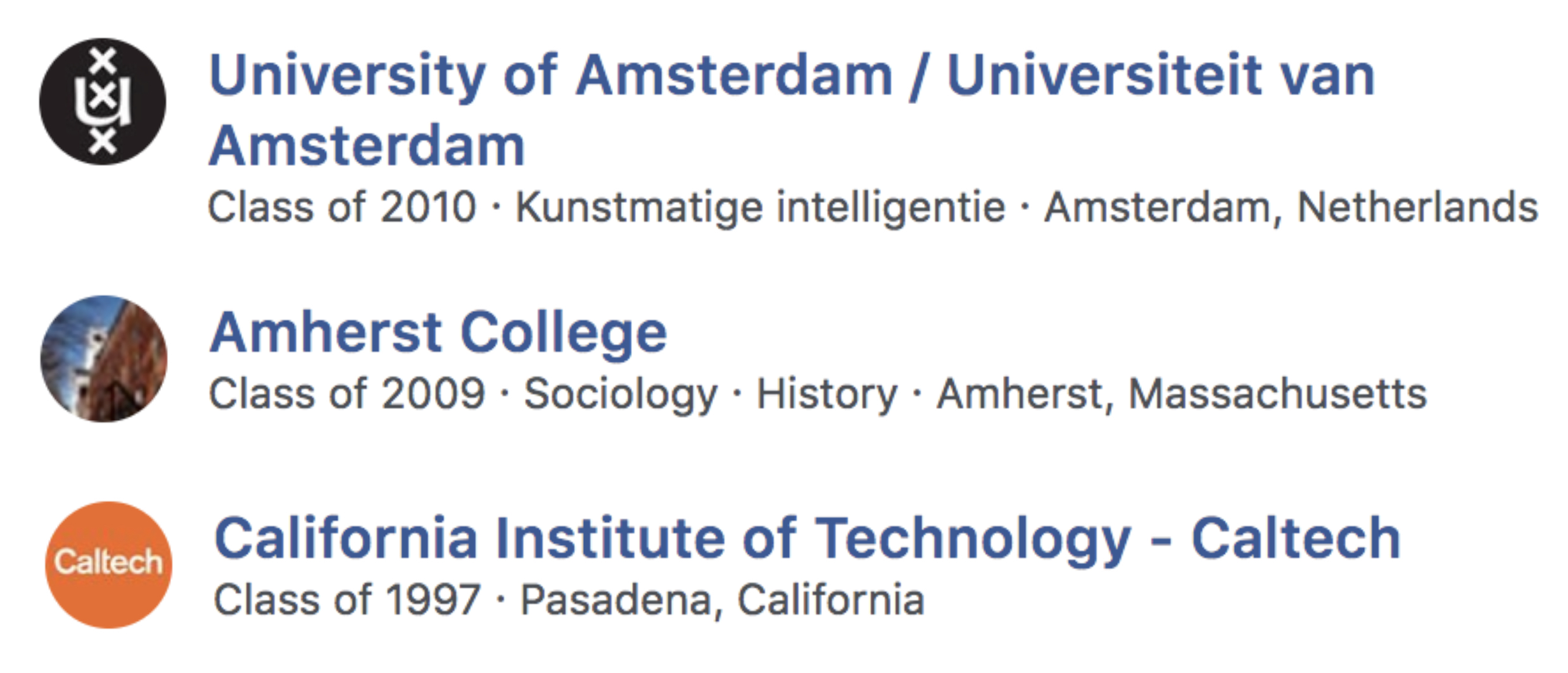}
\caption{
  Examples of self-reported education experiences on Facebook, as provided by the authors.}
\label{fig:example-experience}
\end{figure}

\xhdr{Self-reported demographic information} about age, hometown, and school attendance may optionally be provided by users of the Facebook platform.
For school attendance, this minimally includes the name of the institution of higher education attended by the individual.
Information about the class (graduation year), major (academic specialization) and location of the institution may also be provided.
Example self-reports are shown in Figure \ref{fig:example-experience}.
Self-reports of higher education attendance are assumed to be generally trustworthy,
if they can be resolved against a known U.S. institution of higher education,
and if a sufficient number ($n=10$) of one's Facebook friends
have likewise been identified as attending the same institution.
This preserves 75.4\% of the population, on average, across schools. 
Since our analysis relies in part on the place of residence before going to college,
  we also use self-reported information about one's ``hometown''.
If no hometown was reported, we use the city associated with the most recent high school attended as a proxy.
Of those that report attending a known college, 98\% also report a hometown or high school.

\begin{figure}[t]
  \centering
  \includegraphics[width=0.95\columnwidth]{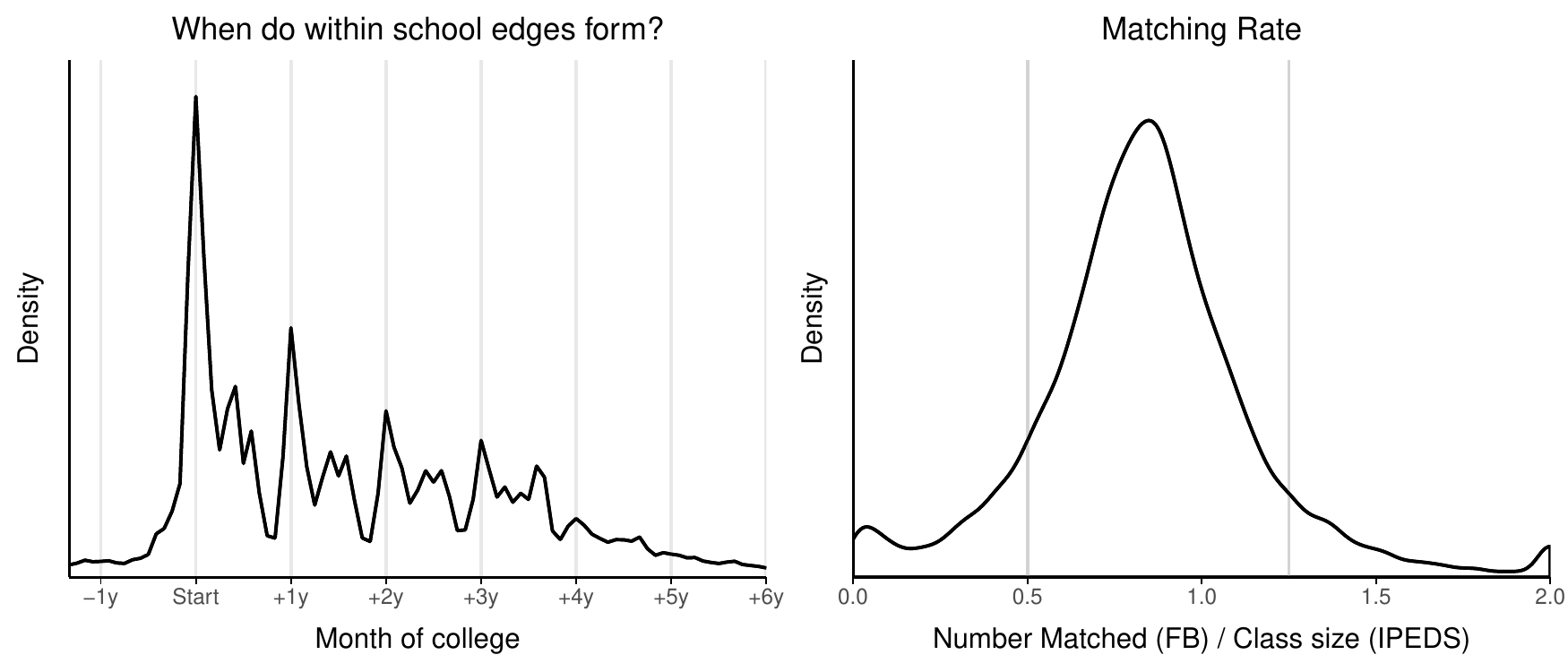}
  \caption{
    \textbf{Left}: Distribution of the timing of new edges (grouped by starting month) within the same college. Shown is the \textit{average} distribution over all schools, aligned by start date. Most new ties are created right around the start of college, with additional spikes at the beginning of each new school year and academic term. \textbf{Right}: Distribution of the matching rate of individual classes. The matching rate is the ratio between the number of people we assigned to a class, and the number of people enrolled at the beginning of the academic school year according to College Scorecard. The x-axis is censored at 2.0. The vertical lines represent the range of 0.5 and 1.25, to which we limit our analysis.}
  \label{fig:data1}
\end{figure}

\xhdr{Facebook friendships} represent our measure of social tie formation. 
Ordinarily, it is difficult to make the assumption that two individuals forming a new Facebook tie have just met.
Existing friends may ``add'' one another after being on the platform for some time,
potentially after seeing Facebook's ``People you may know'' recommendations.
However, in the restricted context of college, this interpretation of a Facebook friendship as indicative of a newly formed tie becomes more viable.
As Figure \ref{fig:data1} (left) shows,
  most Facebook ties with fellow university students are formed during the early days of one's college career.
  Other spikes happen at the beginning of each consecutive academic year and term,
  which are likely driven by the influx of new students into the school and meeting new people when classes start.
These patterns support the assumption that in the setting U.S. higher education the formation of new Facebook ties closely follow the creation of offline social ties.

\xhdr{Institutions of higher education in the United States} are enumerated in the College Scorecard dataset released by the U.S. Department of Education.\footnote{Data retrieved from the College Scorecard website,
          \scriptsize\url{https://ed-public-download.app.cloud.gov/downloads/CollegeScorecard_Raw_Data.zip}.}
We only consider public and private non-profit institutions with an average entry cohort size of at least 100 students.
Additionally, only institutions which appeared in the College Scorecard for at least 4 years during the period 2008--2015 were considered.
For schools with some missing data, we imputed missing covariates as the average of the available years.
The focus of our study is the social networks of undergraduate 4-year higher-education establishments in the United States.
Institutions providing only graduate-level instruction (as of 2015) were removed from our analysis.
All school covariates, including admission and graduation rates, school type and minority status, come from this data.\footnote{
School covariates were constructed as follows.
Admit rate comes from \texttt{ADM\_RATE}.
Graduation rates are based on the \texttt{C150\_4} field,
  which represents the share of students that graduate within 6 years after starting.
The school type (public/private) comes from the \texttt{CONTROL} field.
Some schools are designated as primarily serving minority populations,
  including historically Black colleges and universities (\texttt{HBCU}),
  Hispanic-serving institutions (\texttt{HSI}) and women's-only institutions (\texttt{WOMENS}).
Schools with a religious affiliation are labeled with \texttt{RELAFFIL}.
}
Additionally, we extract a number of school characteristics from the IPEDS system maintained by the NCES.
Yearly class sizes and composition come from the Fall Enrollment dataset.\footnote{Retrieved from 
  \scriptsize\url{https://nces.ed.gov/ipeds/datacenter/data/EF20**.zip}}
We classify schools as dormitory schools, if the school has the capacity to house at least 50\% of its students,
  according to the IC dataset,\footnote{Retrieved from 
  \scriptsize\url{https://nces.ed.gov/ipeds/datacenter/data/IC20**.zip}} and we classify them as commuter schools otherwise.

Matching of Facebook pages and canonical entities provided in the College Scorecard dataset was done as follows.
First, multiple Facebook pages judged to represent the same institution are combined 
  into a single ``metapage'',
for which a representative page is chosen.
The name of this representative page was matched against names of canonical higher education institutions provided in the College Scorecard dataset.
Before matching, both sets of name strings were normalized using a set of heuristics
(e.g. text was lower cased, ``the university of'' was rewritten as ``university of'', etc.)
and some matches had to be manually fixed.\footnote{
This procedure did not work for all institutions. For example the College of Saint Benedict and Saint John's University are considered
as separate institutions by the College Scorecard dataset, but as a single Facebook Page.
}
Finally, we restricted our analysis to matched institutions with at least 100 self-reported students,
  as well as institutions where 50\% or more of the self-reports come from U.S. users.
This procedure produces a total of 1,342 U.S. institutions of higher education, which will get further reduced in the next step.

\xhdr{Entry-year cohorts} (or the ``class of'') are an important organizing structure for social life during college.
Only 18\% of self-reports contain the graduating class so we assign people to classes with the following approach.
First, we assume that students who report their class year are otherwise similar to those who do not.
To support this assumption, we report summary statistics on the two populations in Table \ref{tab:tab_comparison}.
All statistics are computed at the time when the student started college, according to our year assignment.
Students who do not provide a starting year are on average one year younger and are slightly more likely to be male.
Though their profile age does not differ much, they did have 6\% fewer Facebook friends when they started college which could be due to them being less active on Facebook at that time.
With this assumption, we treat the assignment process as a multi-class classification problem,
  with the people who report their year as labeled training examples, and those who do not as instances to be labeled.
We combine features pertaining to the individual, the timing of their friending activity, and of the institution.
We include the institution type as different institutions have different student populations.
For example, age is more strongly correlated with starting class for smaller private schools, so excluding school type would lead to a higher error rate for larger public schools.
For befriending, we include the month with the most new within-school friendships,
  and the 5th percentile (in terms of time) of new friendships forming.
We hypothesize that both of these features are correlated with starting to attend a college.

\begin{table}[t]
\centering 
\begin{tabular}{lrr} 
\\ Statistic    & Year Provided   & Year Imputed \\ 
\hline \\[-1.8ex] 
  Age (years)         &   19.2 (2.6)  &   18.2 (0.9) \\ 
  Profile age (years) &  2.96 (1.93)  & 2.87 (1.74) \\ 
  Number of friends   & 624.4 (568.8) & 589.3 (461.8) \\ 
  Share male          &  0.414 (0.50) & 0.443 (0.50) \\
\end{tabular} 
  \caption{
  \textbf{Comparison of summary statistics of the population of students that provided a starting year and those for whom a year was imputed.} Reported are the means and standard deviations across years. The statistics are computed at the time when school started. Students who do not provide a starting year are on average one year younger, have 6\% fewer Facebook friends, and are slightly more likely to be male.}
  \label{tab:tab_comparison}
\end{table}

We use the generally well-performing method of gradient boosting machines, a variant of Random Forest.
The target classes include the entry-year classes starting in 2008-2015,
and years 2007 and 2016 as bookend classes to capture those outside our range.
The average year-specific cross-validated training accuracy is 69.5\%.  
We assign those who do not self-report a class to the most likely predicted class,
  but only when the predicted class probability for that individual is larger than 0.75.
We do not include individuals whom we cannot assign to a specific class.
For every class, we find the week with the most new within-class friendships forming and 
take that to be the start of the school year.
We manually inspected these values for a number of schools,
and they were generally within one to two weeks of the actual start of the academic school year,
as taken from the institution's official website.

We construct a school-level statistic to measure the rate of participation in Greek life.
First, we identify Facebook groups where at least 75\% of the members are from the same college.
We label as ``Greek groups'' those groups with Greek letters in the name, but excluding known religious and honor societies.
Then we count the share of people in each college that are a member of a Greek group,
  as a share of those that are a member of any college-related group (to make the measure less confounded by Facebook usage).

The IPEDS data provides for each entry year cohort, the number of full-time, first-time undergraduates.
We compare this to the number of people in our dataset that we assigned to that year, to get a match rate for each class.
As can be seen in Figure \ref{fig:data1} (right), the majority of classes in our constructed dataset
  are slightly smaller in size to those according to the College Scorecard data.
We consider a class in our study to have a satisfactory match if we could assign between 0.5 and 1.25 times the average entry cohort size (as derived from the College Scorecard dataset).
Classes that fall outside of this range are excluded from our analysis.\footnote{As a robustness check, we also performed the analyses in this paper using alternative confidence thresholds. Doing so did not meaningfully change any results.}
After this step, 7,660 entry-year classes remain from a total of 1,159 schools, as for some schools there are no classes for which the matching rate is between 0.5 and 1.25.

Given that our sample is biased towards people who use Facebook,
  one may question how representative the people in our dataset are, when compared to the actual student populations.
Both with respect to the gender of students and whether students are predominantly from within-state,
  the classes in our data have a similar composition as those as reported by IPEDS.\footnote{
  Actual class compositions computed based on the Fall-Enrollment reports as released by the IPEDS.
  For share males, the correlation coefficient is $\rho = 0.947$.
  For share in-state, the correlation coefficient is $\rho = 0.843$.}

\section{Data description}

The resulting dataset contains 232.5 million within-cohort Facebook friendship ties between 7.6 million users assigned to 7,660 entry year cohorts in 1,159 U.S. institutions of higher education.
A further 320.1 million edges occur between cohorts but at the same institution. 756.5 million edges connect users assigned to different institutions in our study.
Of the 1,159 institutions in our sample, according to the Carnegie Classification\footnote{
\citep{carnegie15}, retrieved from \url{http://carnegieclassifications.iu.edu}}, 64 are Historically Black Colleges and Universities, 31 are women’s-only institutions, 125 are classified as Hispanic-Serving Institutions, and 221 are undergraduate-only institutions.

\begin{figure}[t]
  \centering
  \includegraphics[width=\columnwidth]{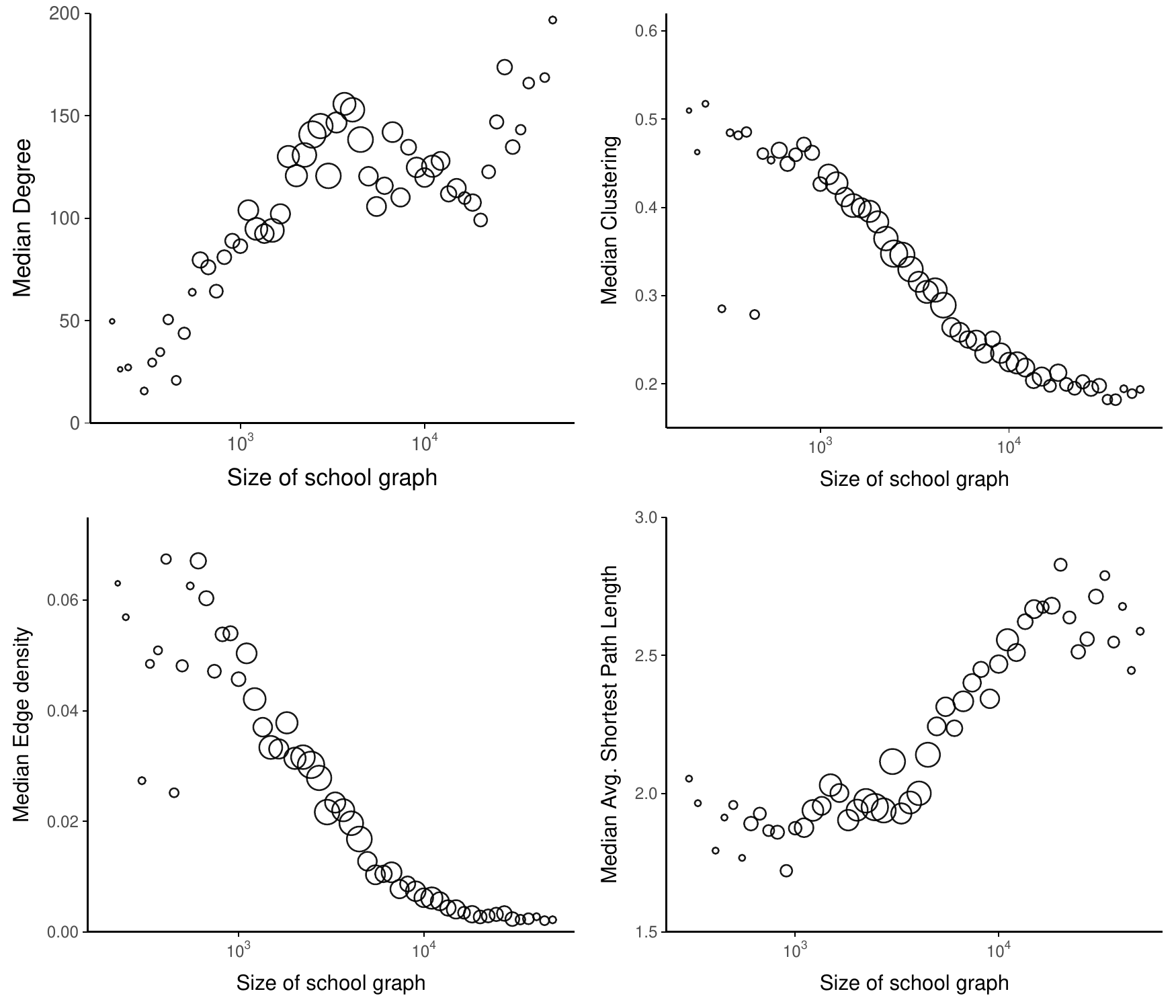}
\caption{
  Structural features of the school graphs, as compared to graph size.
  Graphs are binned logarithmically by the number of nodes in the graph, and for each group we computed the median of the statistic.
  Clockwise from the top-left: degree, local clustering coefficient, average shortest path length, and edge density.
  Average degree increases by graph size, up to about 150.
  Edge density and clustering decrease by graph size.
  Average path length increases by graph size.
  }
\label{fig:data2}
\end{figure}

The 1,159 school graphs vary by size and density,
  both of which have a mechanical effect on other structural measures.
We plot various aggregate network statistics as compared to the size of the school graph in Figure \ref{fig:data2}.
Networks are logarithmically binned by network size.
Note that this comparison is similar to the one done for the FB100 data \citep{jacobs15}.
People in larger school graphs have a higher average degree.
  However, this trend stops at an average degree of 150,
  which is in line with prior observations of there being a soft upper limit to the size of social networks \citep{hill03}.
Though not shown, the degree distributions are not as skewed as is commonly found in social networks.
Since degree grows slower than graph size, edge density also decreases by graph size.
Similarly the average local clustering coefficient also decreases in larger networks, as your friends are less likely to be friends.
This is true, even when accounting for the decreased density.
This relationship between network size and clustering coefficient has been identified in
other empirical social networks as well \citep{leskovec08a,leskovec08b,jacobs15}.
The average shortest path length increases approximately by $O(\log n)$ with the size of the graph $n$, as is expected from random graph theory.

\begin{figure}
  \centering
  \includegraphics[width=\columnwidth]{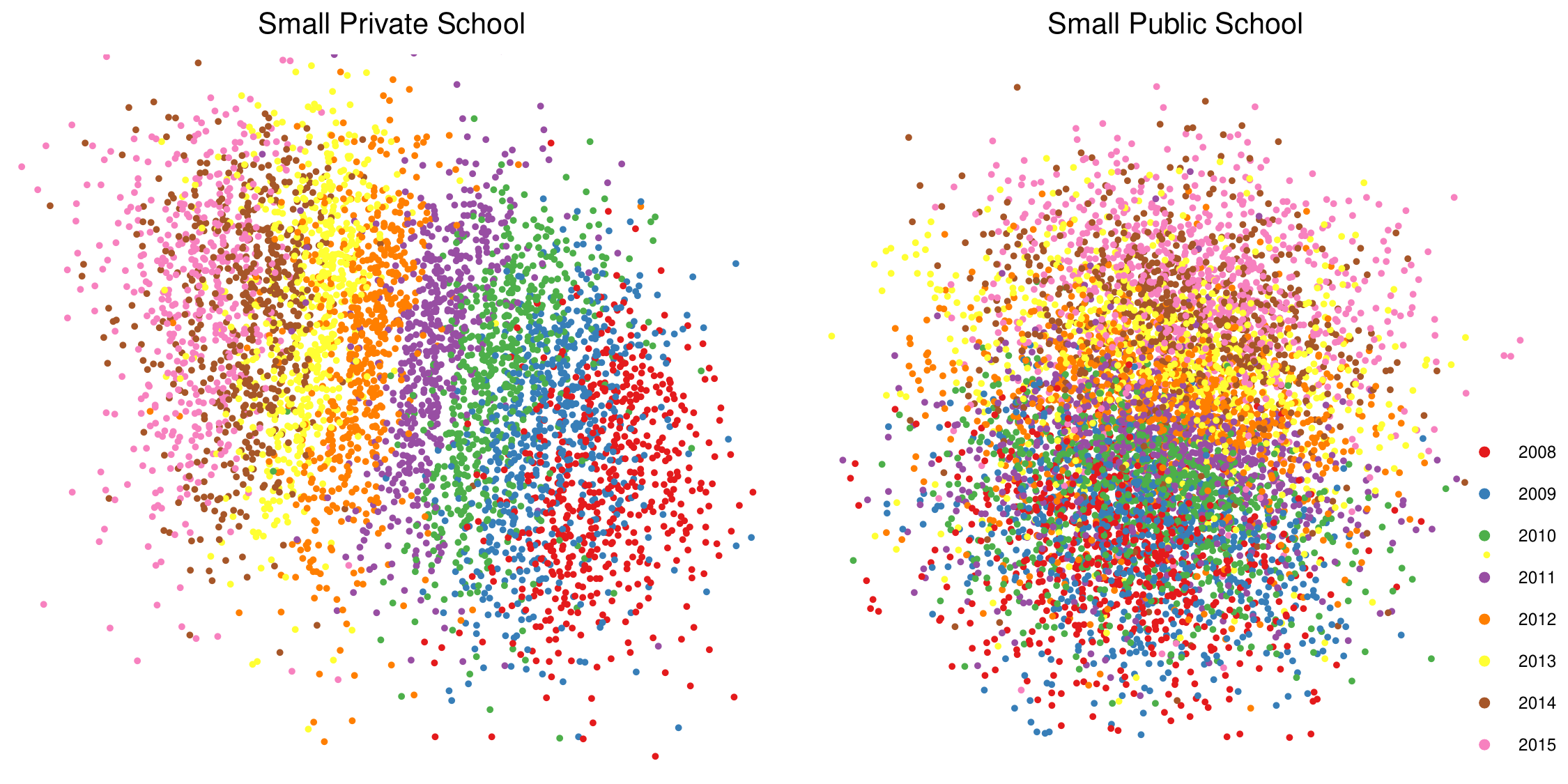}
  \caption{
    The school networks of a small private school and a similarly sized public one.
    Only nodes are shown, colored by each individual's entry-year class.
    The nodes are positioned by the Fruchterman–Reingold projection of their connections.
    The classes of the public school are more mixed than the ones of the private one.
  }
  \label{fig:public_private}
\end{figure}

Schools also display significant heterogeneity with respect to how much mixing there is between students of different entry-year classes.
For example, students in a specific small private liberal-arts college have a higher proportion of within-class friendships than students at a similarly sized public school.
This is shown visually in Figure \ref{fig:public_private}, where both schools' networks are presented
  with a Fruchterman–Reingold projection \cite{fruchterman91} of the connections.
Both schools have a similar structure where adjacent class years are placed next to each other,
  but the students at the private school are more clearly separated by year.
More formally, the modularity score $Q$ \cite{newman04} for the labeled partition by year is 0.309 for the private school and 0.133 for for the public school.

\section{Graph similarity as separability}
  
We want to construct a more formal way to represent structural similarities between networks. 
In brief, we look at how hard it is to tell apart sampled ego-graphs from two schools.
To do so, we train a random forest classifier on each pair of schools,
  and interpret this pairwise classification accuracy as distance.
Now we go over these steps in detail.
 
We represent each network by samples of its ego-graphs.
For each entry-year class graph, we construct $n=250$ ego-graphs by sampling (with replacement) a seed node each time and
  taking its direct (1-hop) neighbors (within the school) and the edges between them.
We only consider edges (and thus neighbors) that existed 4 years after the start of college,
  so the ego-graph approximates what a student's immediate social network looked like when they left college.
This removes the bias where individuals from earlier years have had more time to grow their Facebook networks during college and after.
However, it is a coarse approximation, since less than half of U.S. college students actually graduate within four years.\footnote{NCES Digest of Education Statistics 2017, Table 326.10.}
For each ego-graph we construct the following features:

\begin{itemize}
  \item size -- the number of nodes
  \item the mean and variance of the degree distribution
  \item edge density -- the share of node pairs that are connected
  \item share of nodes from the same year as the ego node
  \item degree assortivity \citep{newman03}
  \item algebraic connectivity \citep{fiedler73}
  \item average clustering coefficient \citep{watts98}
  \item modularity of the modularity-maximizing partition \citep{clauset04}
  \item eigenvector and betweenness centralization \citep{freeman78}
  \item number of connected components of $k$-Cores \citep{bollobas01} and $k$-Brace \citep{ugander12} for $k \in \{8, 16\}$
\end{itemize}

The ego-graphs represent samples from the network that the seed node was drawn from.
This approach is similar in spirit to the NetSimile framework \citep{berlingerio12},
  in that we characterize graphs by computing statistics over sampled subgraphs. 
However, we use a different set of statistics, relevant to our particular domain,
  and rather than aggregating them, we use them as features in a classifier.
We measure similarity between two graphs as the difficulty of telling which one produced what sample.
If a classifier cannot separate two groups of samples,
  so if the predictive accuracy is low,
  we think of the graphs that produced them as similar.
This way of measuring similarity with a prediction task has some precedents in social science, 
  see for example work by Gentzkow, Shapiro and Toddy (\citeyear{gentzkow16}) and Bertrand and Kamenica (\citeyear{bertrand18}).

\begin{figure}
  \centering
  \includegraphics[width=\columnwidth,keepaspectratio]{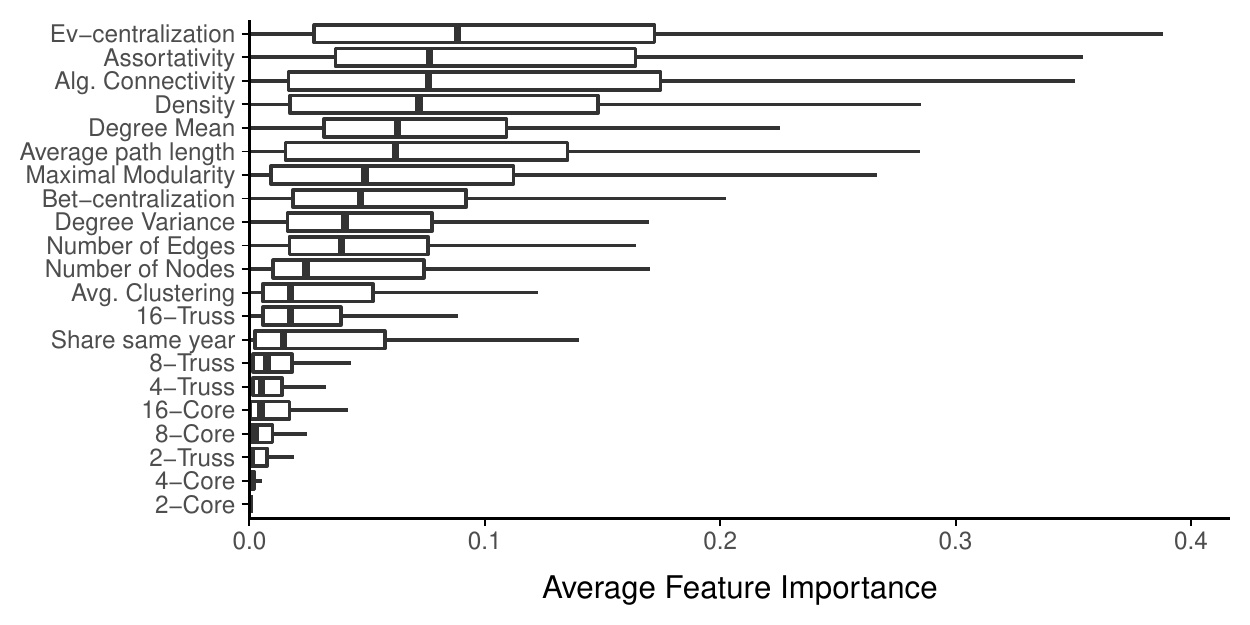}
  \caption{
    Aggregate feature importance for the pairwise ego-graph classifiers, sorted by the median.
    The features that are most often the most distinguishing are eigenvector centralization, degree assortativity, and algebraic connectivity.
  }
  \label{fig:feature_importance}
\end{figure}

For each pair of classes (starting cohorts), as well as for each pair of schools,
  we train a separate random forest model where
  the class/school is the binary outcome label,
  the ego-graphs are the training examples,
  and the listed structural characteristics are the features.
The features thus play a different role to distinguish each pair of classes. 
Because features may be correlated with one another, the datasets generated for each school pair may suffer from the problem of multicollinearity. This is an important reason for our choice of a random forest, a non-linear model which typically does well in settings that contain features of varying quality and with potentially redundant information.
One potential limitation of this approach is that the distance measure is sensitive to the most characteristic statistic.
If one feature is significantly different, but the others are all similar,
  then the two graphs will seem very distinguishable,
  even though they could otherwise be considered very similar.
This could be an issue with size, for example, as
  we know that students at larger schools have a higher average degree and share other characteristics (Figure \ref{fig:data2}).
To investigate whether this happens, we plot the average feature importance for each feature,
  over all trained models, in Figure \ref{fig:feature_importance}.
Edge density, which is correlated with graph size, is indeed among the most important features, but not uniquely so.
Eigenvector centralization and degree assortivity are almost as important on average. 
Clustering, which is more correlated with size than density, is close to the bottom of the list. 

For every pair of graphs (schools or classes),
  the cross-validated area under the curve (AUC) of the model for that pair is then taken to be a measure of distance between the graphs.\footnote{Note that our method is technically not a distance function,
  as it is not symmetric due to the randomness involved in making the cross-validation sets.}  
If the AUC is low (close to 0.5), then the two graphs cannot be easily distinguished, and are thus considered to be similar.
If the AUC is high (close to 1.0), then the two graphs are very easy to distinguish, and thus different.

\section{Graph similarity by school covariates}

\begin{figure}
  \centering
  \includegraphics[width=\columnwidth]{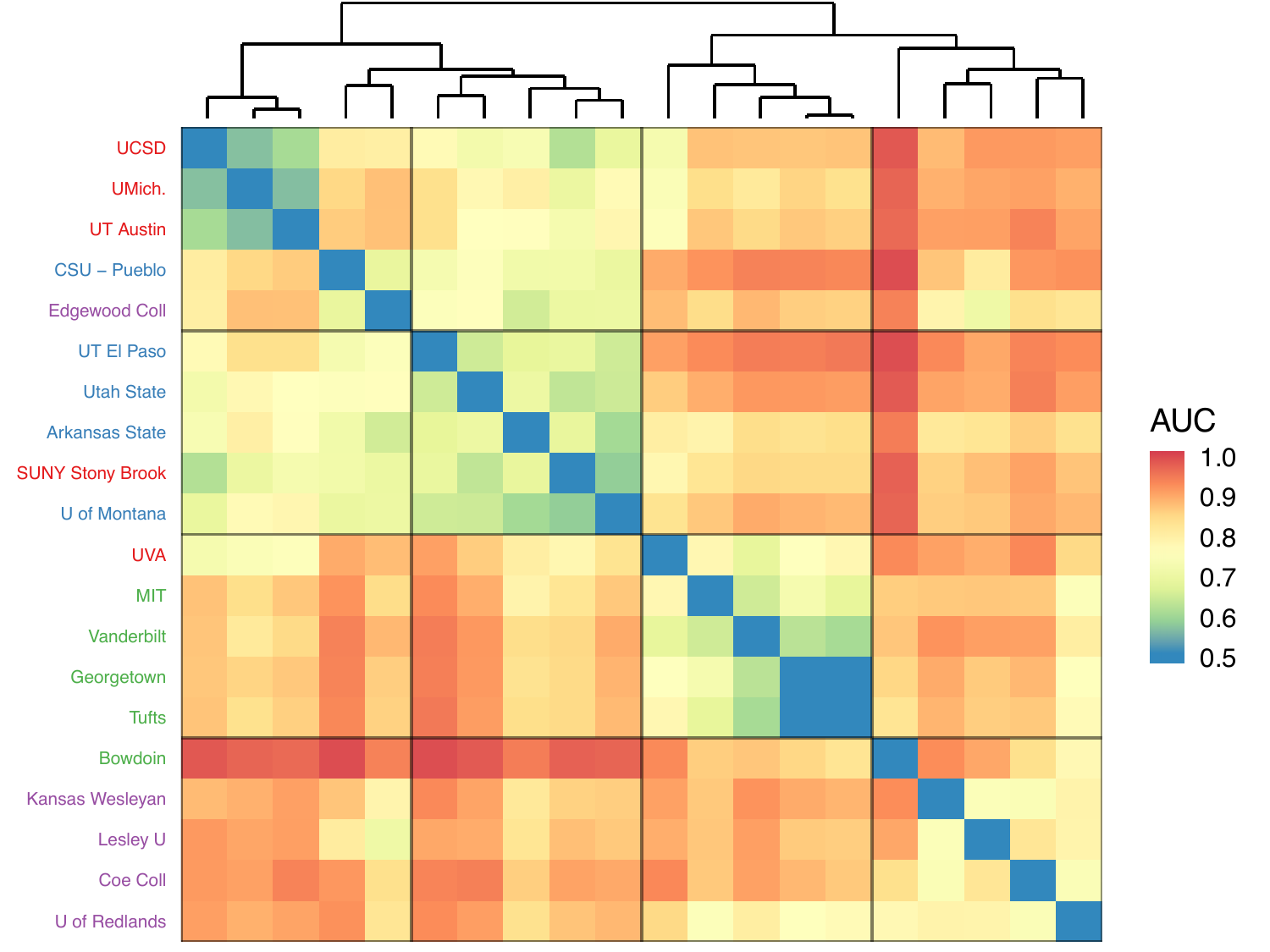}
  \caption{
     Cross-validated AUC of classifiers separating the classes of 2011 for a sample of colleges.
     A high AUC indicates that the graphs are easier to distinguish, and less similar.
     The schools are grouped into four types: public/low-admit (red), public/high-admit (blue), private/low-admit (green), and private/high-admit (purple).
     The rows and columns are identical, and sorted by hierarchical clustering (complete linkage).
     The clustering mostly puts school types together, with some exceptions.
     }
  \label{fig:heatmap-classes}
\end{figure}

A sample of the resulting pairwise AUC values are displayed in Figure \ref{fig:heatmap-classes}.
We sampled five schools from four groups that cross the public/private and low/high admit rate categories (indicated by color) for a total of 20 schools.
For each combination of schools, we plot the AUC for the class graph of 2011, and sort them by by hierarchical clustering.
The schools are mostly organized by similar types, with some exceptions.
The University of Virginia (UVA) is more similar to the selective private schools.
Ego-graphs from Georgetown University and Tufts University are very hard to distinguish.
Bowdoin College has the most uniquely structured ego-graphs.
High-admit private schools have a lower internal similarity than the other groupings,
  potentially due to more heterogeneity in this group.

\begin{figure}[t]
  \centering
  \includegraphics[width=\columnwidth]{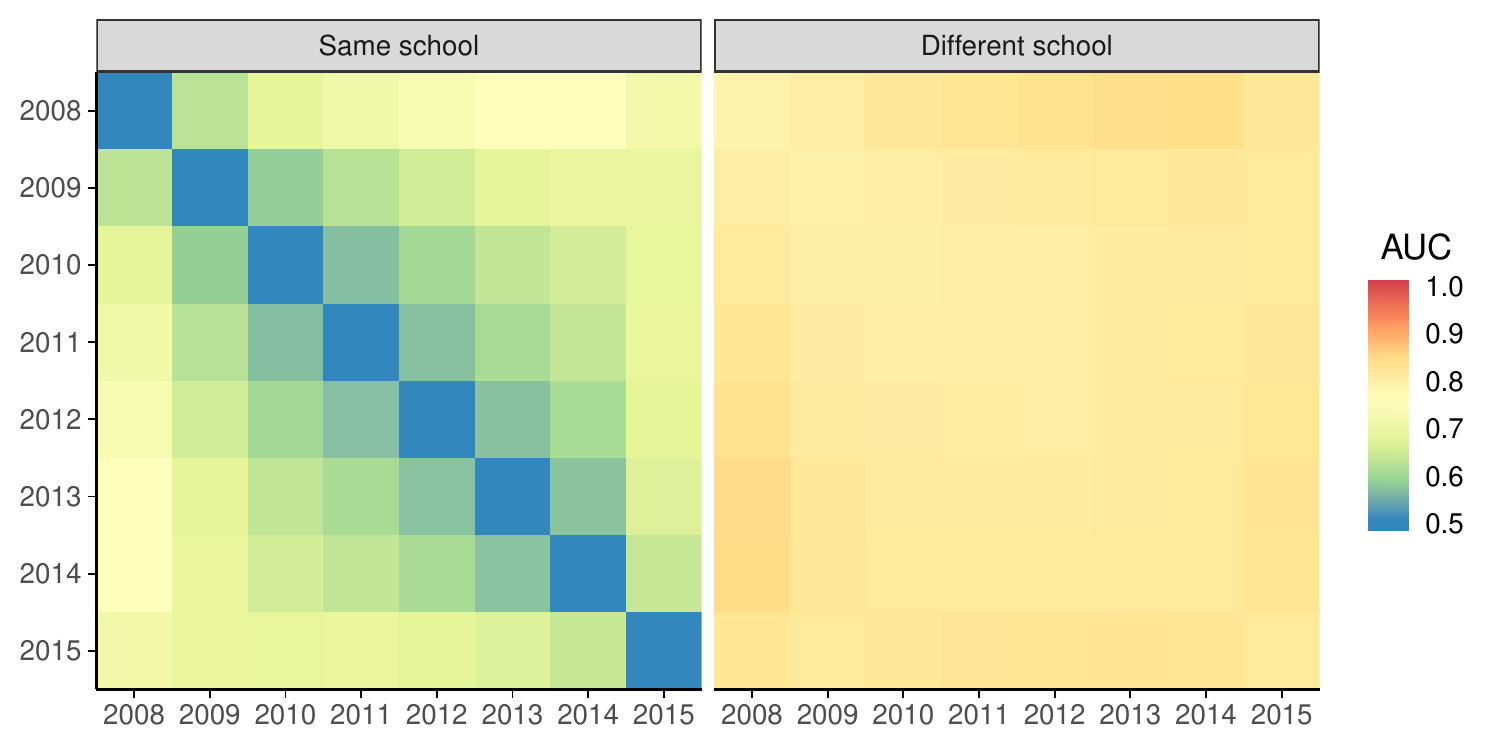}
  \caption{
     Average cross-validated AUC of classifiers separating entry-year classes.
     A high AUC indicates that the graphs are easier to distinguish, and less similar.
     Classes from the same school are harder to distinguish than those from different schools.
     Within the same school, classes that are closer together in time are harder to distinguish than those further in time.
  }
  \label{fig:between-schools}
\end{figure}

As a sanity check to confirm the difference between classes within and across schools,
  we plot the population-wide average AUC's of classes from both groups in Figure \ref{fig:between-schools}.
As expected, classes from the same school (left panel) are on average harder to distinguish than those taken from different schools (right panel).
This is unsurprising, as sampled ego-graphs from the same school are going to partially overlap and will thus share structural similarities.
Within the same school, classes from years that are closer together are harder to distinguish than those further away in time,
  suggesting that network characteristics of schools graphs drift over time. This pattern is not evident in the right panel.
The mechanism behind this finding is unclear, however. It may be that we are witnessing a slow change in social behavior across successive cohorts of students at the same institution. Alternatively, cohorts many years apart may use the Facebook platform in different ways, given the evolution of the website's design over time.

\begin{figure}
  \centering
  \includegraphics[width=\columnwidth]{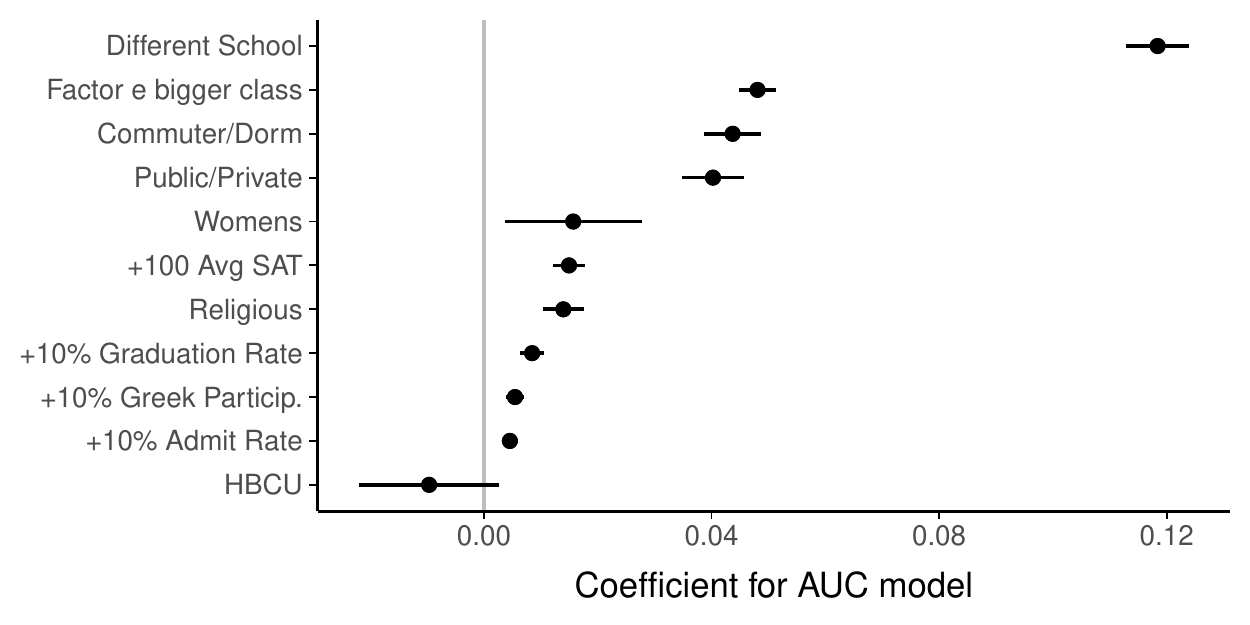}
  \caption{
     Coefficients of AUC regression. Data points are individual random forest models, the AUC is the dependent variable,
     and the \textit{differences} between the two schools are the independents variables. The intercept of the model is 0.5.
     The standard errors are clustered by school.
  }
  \label{fig:auc}
\end{figure}

\begin{figure*}[ht]
  \centering
  \includegraphics[width=\textwidth]{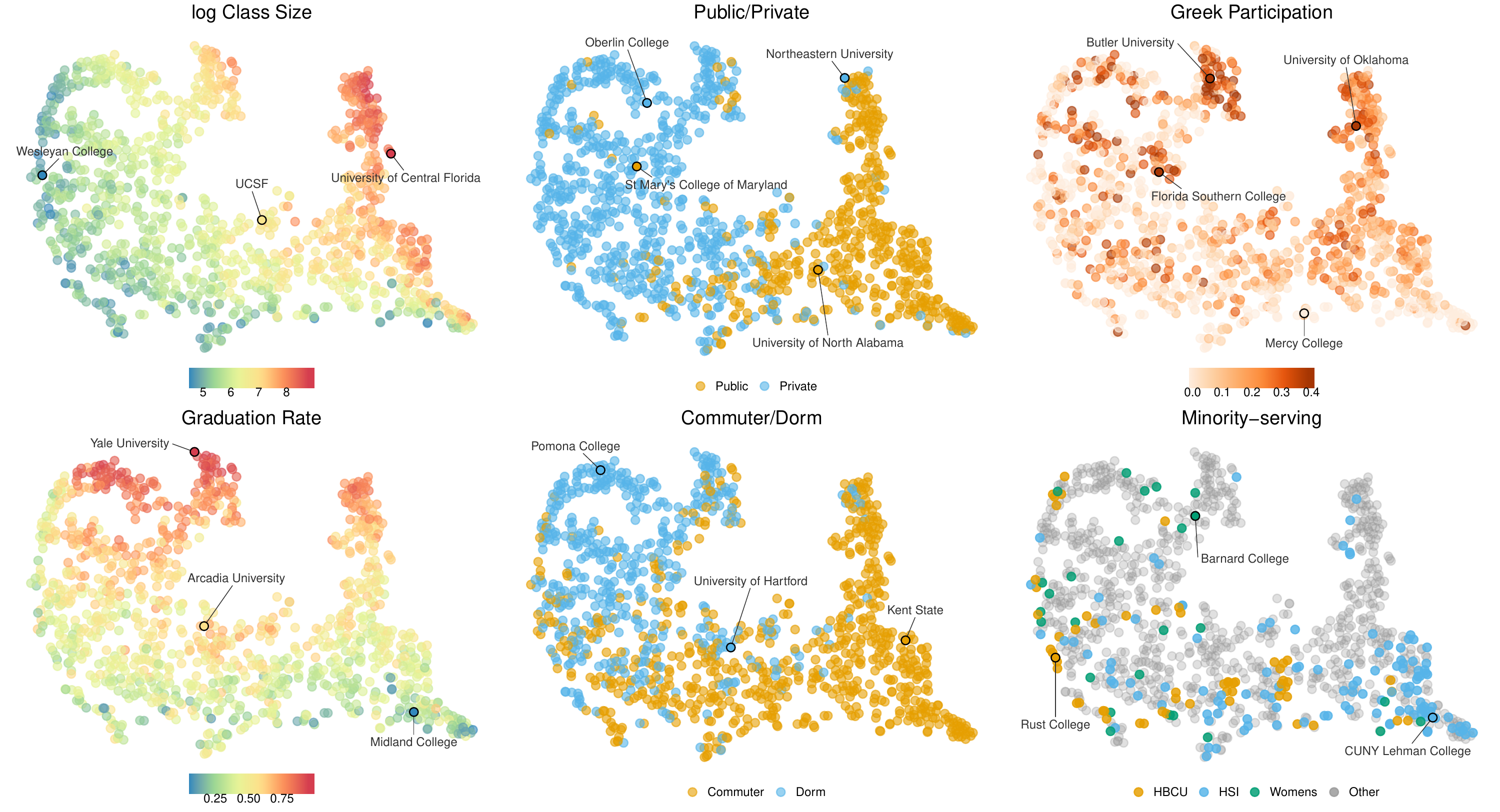}
  \caption{
    t-SNE projection of the average pairwise AUC matrix, colored by various school covariates.
    The main orientation is by class size (horizontal) and by selectivity (vertical).
    The distinctions between both public and private schools, and commuter and dormitory schools, are both clearly visible.
    Schools serving minority populations are not particularly clustered.
  }
  \label{fig:tsne}
\end{figure*}

Next, we look at the average distance between \textit{schools}, rather than classes.
Now, we bin together all ego-graphs taken from the same school,
  and compute the average cross-validated test AUC for every pair of schools. 
We project this pairwise distance matrix to two dimensions with t-SNE~\footnote{We use the \texttt{Rtsne} R library, with perplexity = 40, $\theta=0.2$, although the visual result is similar under different parameters.} \citep{maaten08}.
The resulting projection is shown Figure \ref{fig:tsne}, highlighting different covariates of the individual schools.

A few patterns are immediately apparent.
The schools are distributed by class size on the horizontal axis.
The vertical axis similarly corresponds to selectivity, as displayed by the distribution of schools by graduation rate and admit rate (not shown).
It makes sense that these are the major organizing factors.
Students at bigger schools have a higher degree (Figure \ref{fig:data2}), and a number of other network statistics move accordingly.
Similarly, selectivity is correlated with graduation rates, which affect the number of available people to connect to.
The schools with a low graduation rate are separated into two areas, which corresponds to there being fewer middle-sized schools with a low graduation rate.
The public/private and commuter/dormitory distinctions are both clearly visible as diagonal cuts.
There are some exceptions to this separation, such as Northeastern being placed among mostly public schools and St. Mary's college being placed in the private school region.
The commuter/dormitory distinction is not as clear as the public/private one, which is likely related to it being a discretization of the underlying continuous variable of residential capacity.
The curve on the top left contains mostly small liberal arts schools and the right-most line of schools are all big state schools.
There are pockets of schools with a higher share of Greek life participation, 
  mostly in the very large schools and schools with a very high graduation rate,
  but overall they are not highly clustered in this projection.
Minority-serving institutions are also not particularly clustered, except for HSIs.

Finally, to formalize the relationship between differences in network structure and differences in school types,
  we ran a linear regression with AUC as the dependent variable, and \textit{differences} between schools as the covariates.
For discrete variables, such as whether the school is private, a commuter school, and a minority-serving institution, we construct a binary variable set to 1 when the schools have a different value.
For continuous variables, we take the absolute difference between the schools.
The coefficients of the resulting fit are shown in Figure \ref{fig:auc},
  with the standard errors clustered by schools to account for the repeated involvement of the same school across samples.
The raw difference in AUC between same-school classes (AUC 0.5) and classes from different schools is 30\%.
Accounting for the other school covariates, the baseline difference between these two groups goes to 10.5\%.
Size and the school type are the next biggest distinctive factors.
The average difference between public and private schools is 4.1\%, as is the difference between schools that differ an order of magnitude in log-scale.
Commuter and dormitory schools have a 4.7\% remaining difference.
The differences in other covariates are smaller, but still have a significant effect on how similar the networks are.
An interesting case is the difference between HBCU's and other schools --
  according to our model, the difference between these two is \textit{negative},
  which implies that for these schools being \textit{different} means that they are more similar in terms of network structure.
The $R^2$ for this fit is relatively high at 0.480 (0.707 when adding school fixed effects),
  which indicates that school characteristics indeed capture a substantial part of the variance in predictive accuracy.

\section{Modeling the network structures}

\begin{figure} 
  \centering
  \includegraphics[width=\columnwidth]{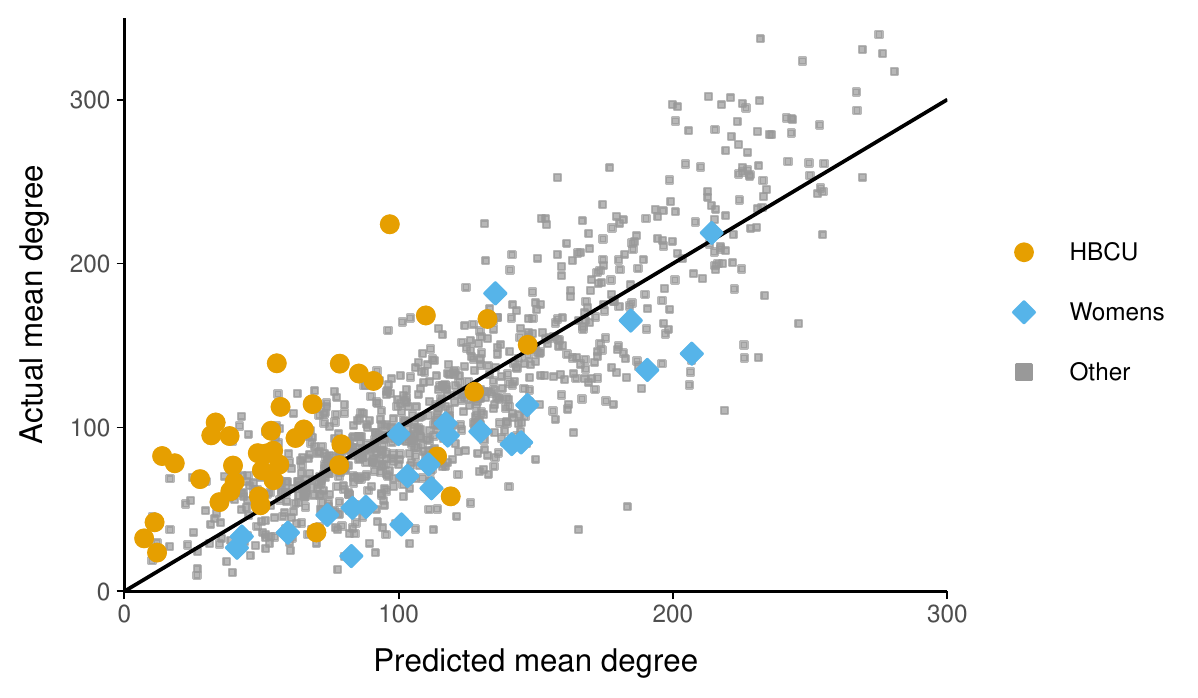}
  \caption{
    Actual average degree by school, as compared to the predicted average degree based on school characteristics.
    The average degree of the 2011 entering classes at colleges is only partly explained by school size, admit rate, public/private, etc. Students at Historically Black Colleges and Universities tend to have higher than predicted average degree based on other school attributes, while women's colleges have lower degree.
  }
  \label{fig:degree_predicted}
\end{figure}

So far we have established that students attending similar kinds of schools
  have consistently similar social networks on Facebook.
In this last section, we directly relate school characteristics to structural outcomes.
We illustrate the need for this with an example.
If one relates the average degree of a student to some school characteristics, 
  the resulting structural deviations are well-explained by the covariates that are left out.
We show this in Figure \ref{fig:degree_predicted}, where we ran a regression on the class graphs of 2011,
  where we relate degree to school covariates (excluding whether the school is a HBCU or women's college) and plot the predicted versus the observed degree.
HBCU's tend to have higher, and women's colleges tend to have lower degree,
  perhaps because cross-gender friendships occur outside of the college.

Next we formalize this insight through a regression model where we include all school covariates we have used to far.
For each entry year class, we take the state of the network four years after school started, to capture the 
network structure when people leave their college due to finishing their undergraduate studies.
This method is admittedly based on an approximation, given that people frequently leave their college earlier (due to dropping out, moving, or transferring),
  or later (due taking more time to finish, staying for graduate school, or post-undergraduate employment).
We compute the following structural features over the year graphs:
  log average degree, the Gini coefficient of the degree, the average clustering coefficient,
  and homophily by year, gender, and hometown.
We chose these outcomes as they are likely to actually affect an individual's experience.
For example, leaving school with more and more diverse social connections may facilitate the finding of work \citep{granovetter73}, whereas a higher inequality in social structure may in turn represent an inequality of opportunity.

We are interested in characterizing the class graphs, but the nodes are embedded in larger school graphs.
Therefore we adjust the network statistics as averages over the nodes in the class graph,
  but compute them over the whole school graph. 
We then run a linear regression model with each structural outcome as the dependent variable,
  and the school characteristics as covariates.
For all but the average degree outcome, we add degree as a covariate (but omit it from the results) to account for the effect that degree has on the other structural measures.
For example, the clustering coefficient is always lower in networks with a lower average degree. 
The units of analysis are social networks by class (entry year cohorts),
    and standard errors are clustered by school.
The results are presented in Table \ref{tab:tab1_post}.
Note that the sample size of 7,168 is lower than the 7,660 classes we mentioned before. This is due to missing values in the school covariates as provided in the Scorecard data, most prominently the admission rate is missing for some schools.\footnote{
The dataset of cohort-level aggregated summary statistics is available for academic use.
Applications for use (a 1-page research proposal) should be sent to \texttt{college\_networks@fb.com}.
}

\begin{table}[t]
\centering 
\footnotesize
\tabcolsep=0.04cm
\begin{tabular}{@{\extracolsep{3pt}}lcccccc} 
\hline \\[-1.8ex] 
\\[-1.8ex] & log    & Gini   & Avg.   & Year & Gender & Town \\ 
\\[-1.8ex] & Degree & Degree & Clust. & H. & H. & H. \\ 
\hline \\[-1.8ex] 
 Class size  & .078$^{*}$ & .029$^{*}$ & $-$.078$^{*}$ & .033$^{*}$ & .016$^{*}$ & .014$^{*}$ \\
$\ \ $ (log)  & (.014) & (.002) & (.002) & (.003) & (.005) & (.001) \\
  Graduation & 1.763$^{*}$ & $-$.168$^{*}$ & $-$.041$^{*}$ & .313$^{*}$ & .154$^{*}$ & $-$.043$^{*}$ \\
 $\ \ $Rate  & (.073) & (.008) & (.010) & (.017) & (.034) & (.006) \\
  Admit  & .025 & .005 & .003 & $-$.001 & .055 & .016$^{*}$ \\
 $\ \ $Rate  & (.051) & (.005) & (.006) & (.010) & (.017) & (.004) \\
  Commuter & $-$.287$^{*}$ & .027$^{*}$ & $-$.015$^{*}$ & $-$.013 & $-$.009 & .005$^{*}$ \\
  & (.021) & (.002) & (.003) & (.004) & (.008) & (.001) \\
  Private & .182$^{*}$ & .011 & .021$^{*}$ & .059$^{*}$ & $-$.0003 & $-$.023$^{*}$ \\
  & (.031) & (.003) & (.004) & (.007) & (.011) & (.002) \\
  Religious & .028 & .003 & .002 & $-$.013 & .037$^{*}$ & $-$.0003 \\
  & (.020) & (.002) & (.003) & (.005) & (.010) & (.001) \\
  Womens & $-$.352$^{*}$ & .004 & .004 & $-$.014 & .800$^{*}$ & $-$.00004 \\
  & (.057) & (.007) & (.010) & (.012) & (.065) & (.002) \\
  HBCU & .315$^{*}$ & .002 & .021 & .118$^{*}$ & $-$.137 & .003 \\
  & (.057) & (.007) & (.008) & (.017) & (.043) & (.004) \\
  Greek & .723$^{*}$ & .079$^{*}$ & .062$^{*}$ & $-$.141$^{*}$ & .138$^{*}$ & $-$.011 \\
  $\ \ $particip.  & (.063) & (.008) & (.008) & (.013) & (.024) & (.004) \\
  log Degree &  & $-$.042$^{*}$ & .014$^{*}$ & $-$.071$^{*}$ & $-$.007 & .003 \\
  &  & (.003) & (.003) & (.003) & (.010) & (.002) \\
  Constant & 2.977$^{*}$ & .498$^{*}$ & .760$^{*}$ & .341$^{*}$ & $-$.185$^{*}$ & $-$.060$^{*}$ \\
  & (.101) & (.014) & (.018) & (.028) & (.047) & (.010) \\
 \hline \\[-1.8ex] 
$N$ & 7,168 & 7,168 & 7,066 & 7,168 & 7,168 & 7,168 \\ 
Adj. R$^{2}$ & .643 & .785 & .786 & .285 & .557 & .573  \\  
\hline 
\hline \\[-1.8ex] 
\textit{Note:}  & \multicolumn{6}{r}{$^{*}$p$<$0.001} \\ 
\end{tabular} 
  \caption{
    \textbf{Linear regression relating post-college network features to school covariates.}
  We fit a separate model with each network statistic as the dependent variable,
    including (log) average degree, Gini degree, average clustering,
    and Newman homophily by year, gender, and hometown  .
  The units of analysis are social networks by class (entry year cohorts),
    and standard errors are clustered by school.
  Network features are computed four years after the estimated start of school.
  Per class, the network features are computed for the people in that class,
    but considering the full school graph at that time (see text).
  We account for (log) average degree in the models where it is not the outcome.
  } 
  \label{tab:tab1_post}
\end{table}

\xhdr{Degree.}
First, we look at the (log transformed) average degree of people from the entry year cohort,
taking into consideration all edges within the school graph.
This can be interpreted as a measure of popularity or gregariousness.
As can be seen in the first column of Table \ref{tab:tab1_post}, there are a number of clear patterns.
The first three effects are likely due to availability.
Students at schools with larger classes have a higher average degree.
However, as was shown in Figure \ref{fig:data2}, this relationship stabilizes for large classes, hence the log-transformation.
Students at schools with lower graduation rates on average have a lower degree, perhaps because of less time during which friendships can form for those students who leave early.
Commuter schools have a lower average degree within, possibly because their students spend less time with each other,
  and them having more active friends outside of schools.
Accounting for this, private schools, HBCU's, and schools with more Greek participation have higher average degrees.
This suggests that attending an institution of these types may lead to greater social connectedness online.
Women's-only institutions have a significantly \textit{lower} degree, perhaps because all opposite-gender connections are made out-of-college at a gender-segregated institution, whereas at least some opposite-gender connections will be created between students at a co-educational institution.

Next we look at inequality in degree, using the Gini coefficient as computed over the degree distribution.
Schools with similar average degrees can nevertheless have very different variances.
For example, students at Haverford College and University of Tennessee both 
  have on average about 200 Facebook friends from their school when they leave college,
  but the Gini coefficient of the former (0.28) is just over half that of the latter (0.48). 
Bigger schools, commuter schools, and schools with more Greek participation have a larger Gini coefficient on average, and thus more inequality in degree.
Schools with a higher graduation rate have more degree equality.
However, from this model it is not clear what mechanisms drive these differences.

\xhdr{Clustering.}
A distinguishing feature of social networks is the extent to which edges are clustered \citep{watts98}.
We measure clustering using the standard average clustering coefficient of nodes of a certain class,
  but taking into account edges to and between all nodes in the school graph. 
The intercept indicates that the population-wide average of the clustering coefficient of the class graphs is very high.
Schools with bigger classes have lower clustering, even when accounting for average degree.
Private schools and schools with more Greek life have more clustering, 
  as these kinds of schools likely have more on campus meeting points.
For the three non-homophily outcomes, the $R^2$ values are quite high,
  again indicating that school characteristics explain a lot of the variance in the outcomes.

\xhdr{Homophily.}
Social networks are often characterized by homophily,
  the tendency to connect to similar others \citep{mcpherson01}. 
We measure homophily across three dimensions: year, gender, and hometown.
For gender and hometown homophily, we use a version of Newman's assortativity coefficient $H$ \citep{newman03}.
However, we again only consider edges involving at least one person from a specific class.\footnote{
For example, in our measure the term $e_{ii}$ from the formula
$H=\frac{\sum_i e_{ii} - \sum_i a_i^2}{1 - \sum_i a_i^2}$ refers to the number of edges coming
from people in a class, where both connected nodes are in group $i$, normalized by the total
number of edges containing people from that class.}
This measure is defined between -1 (only connect with others that are different) and 1.
In practice, the actual range of the measure depends on the relative size and degree distributions of the subgroups, which makes it hard to compare between different networks \citep{cinelli19}.
Luckily for our use case, most compared groups are of similar size, except for some bigger home towns and some schools with highly skewed gender ratios.
Our adjusted version of $H$ does not work for year homophily, as the out-group share in our embedded class graphs would always be 0.
Therefore, for year homophily we simply count the share of edges
  between people of the same entry year, which is defined between 0 and 1.

Year homophily is high across the board, and is highest in bigger schools with a higher graduation rate, as well as in private schools and HBCU's.
Greek life is correlated with more mixing across years, which is intuitive as fraternities and sororities by their very structure offer opportunities for cross-cohort mixing.
Commuter schools also see more mixing across years, which could be because of a more flexible progression throughout school, more pre-college ties, or less dorm life.
In contrast to year homophily, gender homophily is relatively low.
Schools with more Greek life participation have a relatively high degree of gender homophily,
  presumably as there is more same-gender contact.
%HBCU's are an outlier, having distinctly \textit{more} mixing between genders.
Finally, homophily by hometown is also low, an outcome of the large cardinality of the set of potential hometowns.
Hometown homophily is, however, slightly higher in bigger and less selective schools, likely a result of larger groups of students from some hometowns being found at the same institution in these categories.

\section{Limitations and future work}

We hope this paper helps advance social scientists' understanding of the network structure of U.S. colleges and universities by presenting a comprehensive overview of these networks' characteristics. At the same time, there are important limitations that must be acknowledged in discussing this study's contributions. For one, we do not model the formation of social ties, which means we can only speculate on the underlying mechanisms behind the patterns we observe. Furthermore, there are many student characteristics that have been shown to be instrumental
  in network formation in schools,
  including classes that they took, their ethnicity, socio-economic status,
  and, where relevant, their dormitory, which are not studied in this paper.

Our analysis is likely affected by the need to approximate, for those individuals who
  did not specify their years in college, when they started and stopped attending.
Both the accuracy of our year assignment procedure and the four-year graduation rate,
  are correlated with college characteristics,
  which affects how we construct our network data.

More generally, all our analysis applies to social networks on Facebook only.
While some have argued that Facebook networks mirror offline networks, at least structurally \citep{jones13,dunbar15,arnaboldi12,bailey18},
others have observed that the amount of activity on Facebook is correlated with the 
number of Facebook friends \citep{lewis08,viswanath09},
  which would introduce a selection bias.
While we were unable to find public statistics on the usage of Facebook for college-age people for the time frame of our analysis, a survey in 2014 found that 87\% of U.S. people between the ages of 18 and 29 used Facebook \citep{duggan14}.

The networks we presented in this paper are snapshots at the end of college.
However, future work may examine the chronological sequence of edges forming and dissolving \citep{overgoor19}.
Or it might examine the role of pre-college ties in college choice for the formation of college ties.
One could also study the properties of ties originating in college after college ends and
 potentially compare structural measures to known measures of social mobility,
 like the ones prepared by \cite{chetty17}.

Finally, all analysis in this work is observational and
   we do not identify any causal effects.
We leave the mechanisms behind any correlations,
  and the extent to which they are confounded by other factors,
  like Facebook activity, for future work.

\section{Conclusion}

Despite the afore-mentioned limitations, we find it encouraging that it is currently possible to provide a descriptive overview of a class of social networks, as we attempt to do in this study. Universities are a major locus of social interaction, with ties formed in college having reverberations well beyond the campus and into the broader structures of a society. Despite their evident importance, logistical reasons have prevented the comprehensive study of college friendship networks. A particular gap we have sought to address concerns limited knowledge of the network diversity of U.S. higher-education institutions, as well as a lack of insight into variation across years.

In this paper we present a first large-scale analysis of online social networks of U.S. college students across a multi-year timespan. And while some of the conclusions must be tentative, they are also intriguing. We found that, while social networks of different classes within the same institution tend to be structurally similar, same-year networks from different institutions tend to be differentiable. We also showed that these structural differences between social networks formed at different colleges can in part be explained by the attributes of the schools themselves. Larger and public institutions are associated with a smaller number of Facebook friends attending the same college. Graduation rate also correlates with the density of networks. If students are likely to stay and graduate within four years at the school, they are also likely to add more Facebook friends during that period. We also identified a number of structural differences in specific network attributes across school types.

It has been close to thirty years since James Coleman (\citeyearpar{coleman90}) famously identified the glaring gap in the middle of sociology, the lack of ``micro-to-macro'' explanations for how mundane interactions accrete into large scale structures that underpin social life. Social networks have been put forth as a potential solution to this problem, having shown their potential to connect disparate strands of knowledge into a science of society. But for this intellectual project to live up to its full potential, it must be possible for scientists to ask questions across the broad diversity of social networks. Before these questions can be asked, the world of social networks must itself be described -- our goal for this comparatively narrow, yet consequential set of networks. We hope that this description will aid future researchers in understanding the innumerable processes that these networks mediate.

\xhdr{Acknowledgements} We thank Austin Benson, Eduardo Laguna-M{\"u}ggenburg, Johan Ugander, and John Levi Martin for their helpful comments and feedback.

\bibliographystyle{ACM-Reference-Format}
\bibliography{scorecard}


\begin{thebibliography}{55}

%%% the Web address.

\ifx \showCODEN    \undefined \def \showCODEN     #1{\unskip}     \fi
\ifx \showDOI      \undefined \def \showDOI       #1{#1}\fi
\ifx \showISBNx    \undefined \def \showISBNx     #1{\unskip}     \fi
\ifx \showISBNxiii \undefined \def \showISBNxiii  #1{\unskip}     \fi
\ifx \showISSN     \undefined \def \showISSN      #1{\unskip}     \fi
\ifx \showLCCN     \undefined \def \showLCCN      #1{\unskip}     \fi
\ifx \shownote     \undefined \def \shownote      #1{#1}          \fi
\ifx \showarticletitle \undefined \def \showarticletitle #1{#1}   \fi
\ifx \showURL      \undefined \def \showURL       {\relax}        \fi
% invisible to TeX
\providecommand\bibfield[2]{#2}
\providecommand\bibinfo[2]{#2}
\providecommand\natexlab[1]{#1}
\providecommand\showeprint[2][]{arXiv:#2}

\bibitem[\protect\citeauthoryear{Arnaboldi, Conti, Passarella, and
  Pezzoni}{Arnaboldi et~al\mbox{.}}{2012}]%
        {arnaboldi12}
\bibfield{author}{\bibinfo{person}{Valerio Arnaboldi}, \bibinfo{person}{Marco
  Conti}, \bibinfo{person}{Andrea Passarella}, {and} \bibinfo{person}{Fabio
  Pezzoni}.} \bibinfo{year}{2012}\natexlab{}.
\newblock \showarticletitle{Analysis of ego network structure in online social
  networks}. In \bibinfo{booktitle}{\emph{International Confernece on Social
  Computing}}. IEEE, \bibinfo{pages}{31--40}.
\newblock


\bibitem[\protect\citeauthoryear{Arum, Roksa, and Budig}{Arum
  et~al\mbox{.}}{2008}]%
        {arum08}
\bibfield{author}{\bibinfo{person}{Richard Arum}, \bibinfo{person}{Josipa
  Roksa}, {and} \bibinfo{person}{Michelle~J Budig}.}
  \bibinfo{year}{2008}\natexlab{}.
\newblock \showarticletitle{The romance of college attendance: Higher education
  stratification and mate selection}.
\newblock \bibinfo{journal}{\emph{Research in Social Stratification and
  Mobility}} \bibinfo{volume}{26}, \bibinfo{number}{2} (\bibinfo{year}{2008}),
  \bibinfo{pages}{107--121}.
\newblock


\bibitem[\protect\citeauthoryear{Bailey, Cao, Kuchler, Stroebel, and
  Wong}{Bailey et~al\mbox{.}}{2018}]%
        {bailey18}
\bibfield{author}{\bibinfo{person}{Michael Bailey}, \bibinfo{person}{Rachel
  Cao}, \bibinfo{person}{Theresa Kuchler}, \bibinfo{person}{Johannes Stroebel},
  {and} \bibinfo{person}{Arlene Wong}.} \bibinfo{year}{2018}\natexlab{}.
\newblock \showarticletitle{Social Connectedness: Measurement, Determinants,
  and Effects}.
\newblock \bibinfo{journal}{\emph{Journal of Economic Perspectives}}
  \bibinfo{volume}{32}, \bibinfo{number}{3} (\bibinfo{year}{2018}).
\newblock


\bibitem[\protect\citeauthoryear{Berlingerio, Koutra, Eliassi-Rad, and
  Faloutsos}{Berlingerio et~al\mbox{.}}{2012}]%
        {berlingerio12}
\bibfield{author}{\bibinfo{person}{Michele Berlingerio}, \bibinfo{person}{Danai
  Koutra}, \bibinfo{person}{Tina Eliassi-Rad}, {and} \bibinfo{person}{Christos
  Faloutsos}.} \bibinfo{year}{2012}\natexlab{}.
\newblock \showarticletitle{Netsimile: A scalable approach to size-independent
  network similarity}.
\newblock \bibinfo{journal}{\emph{arXiv preprint}} (\bibinfo{year}{2012}).
\newblock


\bibitem[\protect\citeauthoryear{Bertrand and Kamenica}{Bertrand and
  Kamenica}{2018}]%
        {bertrand18}
\bibfield{author}{\bibinfo{person}{Marianne Bertrand} {and}
  \bibinfo{person}{Emir Kamenica}.} \bibinfo{year}{2018}\natexlab{}.
\newblock \bibinfo{booktitle}{\emph{Coming apart? Cultural distances in the
  United States over time}}.
\newblock \bibinfo{type}{{T}echnical {R}eport}. \bibinfo{institution}{National
  Bureau of Economic Research}.
\newblock


\bibitem[\protect\citeauthoryear{Biancani and McFarland}{Biancani and
  McFarland}{2013}]%
        {biancani13}
\bibfield{author}{\bibinfo{person}{Susan Biancani} {and}
  \bibinfo{person}{Daniel~A McFarland}.} \bibinfo{year}{2013}\natexlab{}.
\newblock \showarticletitle{{Social Networks Research in Higher Education}}.
\newblock In \bibinfo{booktitle}{\emph{Higher Education: Handbook of Theory and
  Research}}. \bibinfo{publisher}{Springer}, \bibinfo{pages}{151--215}.
\newblock


\bibitem[\protect\citeauthoryear{Bollob{\'a}s}{Bollob{\'a}s}{2001}]%
        {bollobas01}
\bibfield{author}{\bibinfo{person}{B{\'e}la Bollob{\'a}s}.}
  \bibinfo{year}{2001}\natexlab{}.
\newblock \bibinfo{booktitle}{\emph{Random graphs}}.
\newblock \bibinfo{publisher}{Cambridge studies in advanced mathematics}.
\newblock


\bibitem[\protect\citeauthoryear{Chetty, Friedman, Saez, Turner, and
  Yagan}{Chetty et~al\mbox{.}}{2017}]%
        {chetty17}
\bibfield{author}{\bibinfo{person}{Raj Chetty}, \bibinfo{person}{John~N
  Friedman}, \bibinfo{person}{Emmanuel Saez}, \bibinfo{person}{Nicholas
  Turner}, {and} \bibinfo{person}{Danny Yagan}.}
  \bibinfo{year}{2017}\natexlab{}.
\newblock \bibinfo{booktitle}{\emph{Mobility report cards: The role of colleges
  in intergenerational mobility}}.
\newblock \bibinfo{type}{{T}echnical {R}eport}. \bibinfo{institution}{national
  bureau of economic research}.
\newblock


\bibitem[\protect\citeauthoryear{Cinelli, Peel, Iovanella, and
  Delvenne}{Cinelli et~al\mbox{.}}{2019}]%
        {cinelli19}
\bibfield{author}{\bibinfo{person}{Matteo Cinelli}, \bibinfo{person}{Leto
  Peel}, \bibinfo{person}{Antonio Iovanella}, {and}
  \bibinfo{person}{Jean-Charles Delvenne}.} \bibinfo{year}{2019}\natexlab{}.
\newblock \showarticletitle{Network constraints on the mixing patterns of
  binary node metadata}.
\newblock \bibinfo{journal}{\emph{arXiv preprint arXiv:1908.04588}}
  (\bibinfo{year}{2019}).
\newblock


\bibitem[\protect\citeauthoryear{Clauset, Newman, and Moore}{Clauset
  et~al\mbox{.}}{2004}]%
        {clauset04}
\bibfield{author}{\bibinfo{person}{Aaron Clauset}, \bibinfo{person}{Mark~EJ
  Newman}, {and} \bibinfo{person}{Cristopher Moore}.}
  \bibinfo{year}{2004}\natexlab{}.
\newblock \showarticletitle{Finding community structure in very large
  networks}.
\newblock \bibinfo{journal}{\emph{Physical review E}} \bibinfo{volume}{70},
  \bibinfo{number}{6} (\bibinfo{year}{2004}).
\newblock


\bibitem[\protect\citeauthoryear{Coleman}{Coleman}{1990}]%
        {coleman90}
\bibfield{author}{\bibinfo{person}{James~S Coleman}.}
  \bibinfo{year}{1990}\natexlab{}.
\newblock \bibinfo{booktitle}{\emph{Foundations of social theory}}.
\newblock \bibinfo{publisher}{Harvard University Press}.
\newblock


\bibitem[\protect\citeauthoryear{Currarini, Jackson, and Pin}{Currarini
  et~al\mbox{.}}{2009}]%
        {currarini09}
\bibfield{author}{\bibinfo{person}{Sergio Currarini},
  \bibinfo{person}{Matthew~O Jackson}, {and} \bibinfo{person}{Paolo Pin}.}
  \bibinfo{year}{2009}\natexlab{}.
\newblock \showarticletitle{An economic model of friendship: Homophily,
  minorities, and segregation}.
\newblock \bibinfo{journal}{\emph{Econometrica}} \bibinfo{volume}{77},
  \bibinfo{number}{4} (\bibinfo{year}{2009}), \bibinfo{pages}{1003--1045}.
\newblock


\bibitem[\protect\citeauthoryear{Duggan, Ellison, Lampe, Lenhart, and
  Madden}{Duggan et~al\mbox{.}}{2015}]%
        {duggan14}
\bibfield{author}{\bibinfo{person}{Maeve Duggan}, \bibinfo{person}{Nicole~B
  Ellison}, \bibinfo{person}{Cliff Lampe}, \bibinfo{person}{Amanda Lenhart},
  {and} \bibinfo{person}{Mary Madden}.} \bibinfo{year}{2015}\natexlab{}.
\newblock \bibinfo{booktitle}{\emph{Demographics of key social networking
  platforms}}.
\newblock \bibinfo{type}{{T}echnical {R}eport}. \bibinfo{institution}{Pew
  Research Center}.
\newblock


\bibitem[\protect\citeauthoryear{Dunbar, Arnaboldi, Conti, and
  Passarella}{Dunbar et~al\mbox{.}}{2015}]%
        {dunbar15}
\bibfield{author}{\bibinfo{person}{Robin~IM Dunbar}, \bibinfo{person}{Valerio
  Arnaboldi}, \bibinfo{person}{Marco Conti}, {and} \bibinfo{person}{Andrea
  Passarella}.} \bibinfo{year}{2015}\natexlab{}.
\newblock \showarticletitle{The structure of online social networks mirrors
  those in the offline world}.
\newblock \bibinfo{journal}{\emph{Social Networks}}  \bibinfo{volume}{43}
  (\bibinfo{year}{2015}), \bibinfo{pages}{39--47}.
\newblock


\bibitem[\protect\citeauthoryear{Festinger, Back, and Schachter}{Festinger
  et~al\mbox{.}}{1950}]%
        {festinger50}
\bibfield{author}{\bibinfo{person}{Leon Festinger}, \bibinfo{person}{Kurt~W
  Back}, {and} \bibinfo{person}{Stanley Schachter}.}
  \bibinfo{year}{1950}\natexlab{}.
\newblock \bibinfo{booktitle}{\emph{Social pressures in informal groups: A
  study of human factors in housing}}.
\newblock \bibinfo{publisher}{Stanford University Press}.
\newblock


\bibitem[\protect\citeauthoryear{Fiedler}{Fiedler}{1973}]%
        {fiedler73}
\bibfield{author}{\bibinfo{person}{Miroslav Fiedler}.}
  \bibinfo{year}{1973}\natexlab{}.
\newblock \showarticletitle{Algebraic connectivity of graphs}.
\newblock \bibinfo{journal}{\emph{Czechoslovak mathematical journal}}
  \bibinfo{volume}{23}, \bibinfo{number}{2} (\bibinfo{year}{1973}),
  \bibinfo{pages}{298--305}.
\newblock


\bibitem[\protect\citeauthoryear{Freeman}{Freeman}{1978}]%
        {freeman78}
\bibfield{author}{\bibinfo{person}{Linton~C Freeman}.}
  \bibinfo{year}{1978}\natexlab{}.
\newblock \showarticletitle{Centrality in social networks conceptual
  clarification}.
\newblock \bibinfo{journal}{\emph{Social networks}} \bibinfo{volume}{1},
  \bibinfo{number}{3} (\bibinfo{year}{1978}), \bibinfo{pages}{215--239}.
\newblock


\bibitem[\protect\citeauthoryear{French and Mensh}{French and Mensh}{1948}]%
        {french48}
\bibfield{author}{\bibinfo{person}{Robert~L French} {and}
  \bibinfo{person}{Ivan~N Mensh}.} \bibinfo{year}{1948}\natexlab{}.
\newblock \showarticletitle{Some relationships between interpersonal judgments
  and sociometric status in a college group}.
\newblock \bibinfo{journal}{\emph{Sociometry}} \bibinfo{volume}{11},
  \bibinfo{number}{4} (\bibinfo{year}{1948}), \bibinfo{pages}{335--345}.
\newblock


\bibitem[\protect\citeauthoryear{Fruchterman and Reingold}{Fruchterman and
  Reingold}{1991}]%
        {fruchterman91}
\bibfield{author}{\bibinfo{person}{Thomas~MJ Fruchterman} {and}
  \bibinfo{person}{Edward~M Reingold}.} \bibinfo{year}{1991}\natexlab{}.
\newblock \showarticletitle{Graph drawing by force-directed placement}.
\newblock \bibinfo{journal}{\emph{Software: Practice and experience}}
  \bibinfo{volume}{21}, \bibinfo{number}{11} (\bibinfo{year}{1991}),
  \bibinfo{pages}{1129--1164}.
\newblock


\bibitem[\protect\citeauthoryear{Gentzkow, Shapiro, and Taddy}{Gentzkow
  et~al\mbox{.}}{2016}]%
        {gentzkow16}
\bibfield{author}{\bibinfo{person}{Matthew Gentzkow}, \bibinfo{person}{J~M
  Shapiro}, {and} \bibinfo{person}{M Taddy}.} \bibinfo{year}{2016}\natexlab{}.
\newblock \bibinfo{booktitle}{\emph{{Measuring polarization in high-dimensional
  data: Method and application to congressional speech}}}.
\newblock \bibinfo{type}{{T}echnical {R}eport}. \bibinfo{institution}{National
  Bureau of Economic Research}.
\newblock


\bibitem[\protect\citeauthoryear{Gerber and Cheung}{Gerber and Cheung}{2008}]%
        {gerber08}
\bibfield{author}{\bibinfo{person}{Theodore~P Gerber} {and}
  \bibinfo{person}{Sin~Yi Cheung}.} \bibinfo{year}{2008}\natexlab{}.
\newblock \showarticletitle{Horizontal stratification in postsecondary
  education: Forms, explanations, and implications}.
\newblock \bibinfo{journal}{\emph{Annual Review of Sociology}}
  \bibinfo{volume}{34} (\bibinfo{year}{2008}), \bibinfo{pages}{299--318}.
\newblock


\bibitem[\protect\citeauthoryear{Godley}{Godley}{2008}]%
        {godley08}
\bibfield{author}{\bibinfo{person}{Jenny Godley}.}
  \bibinfo{year}{2008}\natexlab{}.
\newblock \showarticletitle{Preference or propinquity? The relative
  contribution of selection and opportunity to friendship homophily in
  college}.
\newblock \bibinfo{journal}{\emph{Connections}} \bibinfo{volume}{1},
  \bibinfo{number}{1} (\bibinfo{year}{2008}), \bibinfo{pages}{65--80}.
\newblock


\bibitem[\protect\citeauthoryear{Granovetter}{Granovetter}{1973}]%
        {granovetter73}
\bibfield{author}{\bibinfo{person}{M~S Granovetter}.}
  \bibinfo{year}{1973}\natexlab{}.
\newblock \showarticletitle{{The strength of weak ties}}.
\newblock \bibinfo{journal}{\emph{Amer. J. Sociology}} \bibinfo{volume}{78},
  \bibinfo{number}{6} (\bibinfo{year}{1973}), \bibinfo{pages}{1360--1380}.
\newblock


\bibitem[\protect\citeauthoryear{Hanifan}{Hanifan}{1916}]%
        {hanifan16}
\bibfield{author}{\bibinfo{person}{Lyda~J Hanifan}.}
  \bibinfo{year}{1916}\natexlab{}.
\newblock \showarticletitle{The rural school community center}.
\newblock \bibinfo{journal}{\emph{The Annals of the American Academy of
  Political and Social Science}} \bibinfo{volume}{67}, \bibinfo{number}{1}
  (\bibinfo{year}{1916}), \bibinfo{pages}{130--138}.
\newblock


\bibitem[\protect\citeauthoryear{Harris and Udry}{Harris and Udry}{2008}]%
        {harris08}
\bibfield{author}{\bibinfo{person}{Kathleen~Mullan Harris} {and}
  \bibinfo{person}{J~Richard Udry}.} \bibinfo{year}{2008}\natexlab{}.
\newblock \bibinfo{booktitle}{\emph{National longitudinal study of adolescent
  health (add health), 1994-2008}}.
\newblock \bibinfo{publisher}{Inter-university Consortium for Political and
  Social Research}.
\newblock


\bibitem[\protect\citeauthoryear{Hill and Dunbar}{Hill and Dunbar}{2003}]%
        {hill03}
\bibfield{author}{\bibinfo{person}{R~A Hill} {and} \bibinfo{person}{R Dunbar}.}
  \bibinfo{year}{2003}\natexlab{}.
\newblock \showarticletitle{{Social network size in humans}}.
\newblock \bibinfo{journal}{\emph{Human nature}} \bibinfo{volume}{14},
  \bibinfo{number}{1} (\bibinfo{date}{March} \bibinfo{year}{2003}),
  \bibinfo{pages}{53--72}.
\newblock


\bibitem[\protect\citeauthoryear{{Indiana University Center for Postsecondary
  Research}}{{Indiana University Center for Postsecondary Research}}{2015}]%
        {carnegie15}
\bibfield{author}{\bibinfo{person}{{Indiana University Center for Postsecondary
  Research}}.} \bibinfo{year}{2015}\natexlab{}.
\newblock \showarticletitle{The Carnegie Classification of Institutions of
  Higher Education}.
\newblock \bibinfo{journal}{\emph{Bloomington, IN: Author}}
  (\bibinfo{year}{2015}).
\newblock


\bibitem[\protect\citeauthoryear{Jacobs, Way, Ugander, and Clauset}{Jacobs
  et~al\mbox{.}}{2015}]%
        {jacobs15}
\bibfield{author}{\bibinfo{person}{Abigail~Z Jacobs}, \bibinfo{person}{Samuel~F
  Way}, \bibinfo{person}{Johan Ugander}, {and} \bibinfo{person}{Aaron
  Clauset}.} \bibinfo{year}{2015}\natexlab{}.
\newblock \showarticletitle{Assembling thefacebook: Using heterogeneity to
  understand online social network assembly}. In
  \bibinfo{booktitle}{\emph{WebSci}}. ACM, \bibinfo{pages}{18}.
\newblock


\bibitem[\protect\citeauthoryear{Jones, Settle, Bond, Fariss, Marlow, and
  Fowler}{Jones et~al\mbox{.}}{2013}]%
        {jones13}
\bibfield{author}{\bibinfo{person}{Jason~J Jones}, \bibinfo{person}{Jaime~E
  Settle}, \bibinfo{person}{Robert~M Bond}, \bibinfo{person}{Christopher~J
  Fariss}, \bibinfo{person}{Cameron Marlow}, {and} \bibinfo{person}{James~H
  Fowler}.} \bibinfo{year}{2013}\natexlab{}.
\newblock \showarticletitle{Inferring tie strength from online directed
  behavior}.
\newblock \bibinfo{journal}{\emph{PloS one}} \bibinfo{volume}{8},
  \bibinfo{number}{1} (\bibinfo{year}{2013}).
\newblock


\bibitem[\protect\citeauthoryear{Joyner and Kao}{Joyner and Kao}{2000}]%
        {joyner00}
\bibfield{author}{\bibinfo{person}{Kara Joyner} {and} \bibinfo{person}{Grace
  Kao}.} \bibinfo{year}{2000}\natexlab{}.
\newblock \showarticletitle{School racial composition and adolescent racial
  homophily}.
\newblock \bibinfo{journal}{\emph{Social Science Quarterly}}
  (\bibinfo{year}{2000}), \bibinfo{pages}{810--825}.
\newblock


\bibitem[\protect\citeauthoryear{Kossinets and Watts}{Kossinets and
  Watts}{2006}]%
        {kossinets06}
\bibfield{author}{\bibinfo{person}{Gueorgi Kossinets} {and}
  \bibinfo{person}{Duncan~J Watts}.} \bibinfo{year}{2006}\natexlab{}.
\newblock \showarticletitle{Empirical analysis of an evolving social network}.
\newblock \bibinfo{journal}{\emph{Science}} \bibinfo{volume}{311},
  \bibinfo{number}{5757} (\bibinfo{year}{2006}), \bibinfo{pages}{88--90}.
\newblock


\bibitem[\protect\citeauthoryear{Leskovec, Backstrom, Kumar, and
  Tomkins}{Leskovec et~al\mbox{.}}{2008}]%
        {leskovec08a}
\bibfield{author}{\bibinfo{person}{Jure Leskovec}, \bibinfo{person}{Lars
  Backstrom}, \bibinfo{person}{Ravi Kumar}, {and} \bibinfo{person}{Andrew
  Tomkins}.} \bibinfo{year}{2008}\natexlab{}.
\newblock \showarticletitle{Microscopic evolution of social networks}. In
  \bibinfo{booktitle}{\emph{KDD}}. ACM, \bibinfo{pages}{462--470}.
\newblock


\bibitem[\protect\citeauthoryear{Leskovec and Horvitz}{Leskovec and
  Horvitz}{2008}]%
        {leskovec08b}
\bibfield{author}{\bibinfo{person}{Jure Leskovec} {and} \bibinfo{person}{Eric
  Horvitz}.} \bibinfo{year}{2008}\natexlab{}.
\newblock \showarticletitle{Planetary-scale views on a large instant-messaging
  network}. In \bibinfo{booktitle}{\emph{WWW}}. ACM, \bibinfo{pages}{915--924}.
\newblock


\bibitem[\protect\citeauthoryear{Lewis, Gonzalez, and Kaufman}{Lewis
  et~al\mbox{.}}{2012}]%
        {lewis12}
\bibfield{author}{\bibinfo{person}{Kevin Lewis}, \bibinfo{person}{Marco
  Gonzalez}, {and} \bibinfo{person}{Jason Kaufman}.}
  \bibinfo{year}{2012}\natexlab{}.
\newblock \showarticletitle{Social selection and peer influence in an online
  social network}.
\newblock \bibinfo{journal}{\emph{Proceedings of the National Academy of
  Sciences}} \bibinfo{volume}{109}, \bibinfo{number}{1} (\bibinfo{year}{2012}),
  \bibinfo{pages}{68--72}.
\newblock


\bibitem[\protect\citeauthoryear{Lewis, Kaufman, Gonzalez, Wimmer, and
  Christakis}{Lewis et~al\mbox{.}}{2008}]%
        {lewis08}
\bibfield{author}{\bibinfo{person}{Kevin Lewis}, \bibinfo{person}{Jason
  Kaufman}, \bibinfo{person}{Marco Gonzalez}, \bibinfo{person}{Andreas Wimmer},
  {and} \bibinfo{person}{Nicholas~A Christakis}.}
  \bibinfo{year}{2008}\natexlab{}.
\newblock \showarticletitle{{Tastes, ties, and time: A new social network
  dataset using Facebook.com}}.
\newblock \bibinfo{journal}{\emph{Social Networks}} \bibinfo{volume}{30},
  \bibinfo{number}{4} (\bibinfo{date}{Oct.} \bibinfo{year}{2008}),
  \bibinfo{pages}{330--342}.
\newblock


\bibitem[\protect\citeauthoryear{Maaten and Hinton}{Maaten and Hinton}{2008}]%
        {maaten08}
\bibfield{author}{\bibinfo{person}{Laurens van~der Maaten} {and}
  \bibinfo{person}{Geoffrey Hinton}.} \bibinfo{year}{2008}\natexlab{}.
\newblock \showarticletitle{Visualizing data using t-SNE}.
\newblock \bibinfo{journal}{\emph{Journal of machine learning research}}
  \bibinfo{volume}{9}, \bibinfo{number}{Nov} (\bibinfo{year}{2008}),
  \bibinfo{pages}{2579--2605}.
\newblock


\bibitem[\protect\citeauthoryear{Marmaros and Sacerdote}{Marmaros and
  Sacerdote}{2006}]%
        {marmaros06}
\bibfield{author}{\bibinfo{person}{David Marmaros} {and} \bibinfo{person}{Bruce
  Sacerdote}.} \bibinfo{year}{2006}\natexlab{}.
\newblock \showarticletitle{How do friendships form?}
\newblock \bibinfo{journal}{\emph{Quarterly Journal of Economics}}
  \bibinfo{volume}{121}, \bibinfo{number}{1} (\bibinfo{year}{2006}),
  \bibinfo{pages}{79--119}.
\newblock


\bibitem[\protect\citeauthoryear{Mayer and Puller}{Mayer and Puller}{2008}]%
        {mayer08}
\bibfield{author}{\bibinfo{person}{Adalbert Mayer} {and}
  \bibinfo{person}{Steven~L Puller}.} \bibinfo{year}{2008}\natexlab{}.
\newblock \showarticletitle{The old boy (and girl) network: Social network
  formation on university campuses}.
\newblock \bibinfo{journal}{\emph{Journal of Public Economics}}
  \bibinfo{volume}{92}, \bibinfo{number}{1-2} (\bibinfo{year}{2008}),
  \bibinfo{pages}{329--347}.
\newblock


\bibitem[\protect\citeauthoryear{McFarland}{McFarland}{2001}]%
        {mcfarland01}
\bibfield{author}{\bibinfo{person}{Daniel~A McFarland}.}
  \bibinfo{year}{2001}\natexlab{}.
\newblock \showarticletitle{Student resistance: How the formal and informal
  organization of classrooms facilitate everyday forms of student defiance}.
\newblock \bibinfo{journal}{\emph{Amer. J. Sociology}} \bibinfo{volume}{107},
  \bibinfo{number}{3} (\bibinfo{year}{2001}), \bibinfo{pages}{612--678}.
\newblock


\bibitem[\protect\citeauthoryear{McFarland, Moody, Diehl, Smith, and
  Thomas}{McFarland et~al\mbox{.}}{2014}]%
        {mcfarland14}
\bibfield{author}{\bibinfo{person}{Daniel~A McFarland}, \bibinfo{person}{James
  Moody}, \bibinfo{person}{David Diehl}, \bibinfo{person}{Jeffrey~A Smith},
  {and} \bibinfo{person}{Reuben~J Thomas}.} \bibinfo{year}{2014}\natexlab{}.
\newblock \showarticletitle{Network ecology and adolescent social structure}.
\newblock \bibinfo{journal}{\emph{American Sociological Review}}
  \bibinfo{volume}{79}, \bibinfo{number}{6} (\bibinfo{year}{2014}),
  \bibinfo{pages}{1088--1121}.
\newblock


\bibitem[\protect\citeauthoryear{McPherson, Smith-Lovin, and Cook}{McPherson
  et~al\mbox{.}}{2001}]%
        {mcpherson01}
\bibfield{author}{\bibinfo{person}{Miller McPherson}, \bibinfo{person}{Lynn
  Smith-Lovin}, {and} \bibinfo{person}{James~M Cook}.}
  \bibinfo{year}{2001}\natexlab{}.
\newblock \showarticletitle{Birds of a feather: Homophily in social networks}.
\newblock \bibinfo{journal}{\emph{Annual Review of Sociology}}
  \bibinfo{volume}{27}, \bibinfo{number}{1} (\bibinfo{year}{2001}),
  \bibinfo{pages}{415--444}.
\newblock


\bibitem[\protect\citeauthoryear{Moreno}{Moreno}{1934}]%
        {moreno34}
\bibfield{author}{\bibinfo{person}{J~L Moreno}.}
  \bibinfo{year}{1934}\natexlab{}.
\newblock \bibinfo{booktitle}{\emph{{Who Shall Survive: A New Approach to the
  Problem of Human Interrelations}}}.
\newblock \bibinfo{publisher}{Nervous and mental disease monograph series}.
\newblock


\bibitem[\protect\citeauthoryear{Newman}{Newman}{2003}]%
        {newman03}
\bibfield{author}{\bibinfo{person}{Mark~EJ Newman}.}
  \bibinfo{year}{2003}\natexlab{}.
\newblock \showarticletitle{Mixing patterns in networks}.
\newblock \bibinfo{journal}{\emph{Physical Review E}} \bibinfo{volume}{67},
  \bibinfo{number}{2} (\bibinfo{year}{2003}), \bibinfo{pages}{026126}.
\newblock


\bibitem[\protect\citeauthoryear{Newman and Girvan}{Newman and Girvan}{2004}]%
        {newman04}
\bibfield{author}{\bibinfo{person}{Mark~EJ Newman} {and}
  \bibinfo{person}{Michelle Girvan}.} \bibinfo{year}{2004}\natexlab{}.
\newblock \showarticletitle{Finding and evaluating community structure in
  networks}.
\newblock \bibinfo{journal}{\emph{Physical review E}} \bibinfo{volume}{69},
  \bibinfo{number}{2} (\bibinfo{year}{2004}), \bibinfo{pages}{026113}.
\newblock


\bibitem[\protect\citeauthoryear{Overgoor, Benson, and Ugander}{Overgoor
  et~al\mbox{.}}{2019}]%
        {overgoor19}
\bibfield{author}{\bibinfo{person}{Jan Overgoor}, \bibinfo{person}{Austin~R
  Benson}, {and} \bibinfo{person}{Johan Ugander}.}
  \bibinfo{year}{2019}\natexlab{}.
\newblock \showarticletitle{{Choosing to grow a graph: Modeling network
  formation as discrete choice}}. In \bibinfo{booktitle}{\emph{WWW}}. ACM,
  \bibinfo{pages}{1409--1420}.
\newblock


\bibitem[\protect\citeauthoryear{Ryan and Bauman}{Ryan and Bauman}{2016}]%
        {ryan16}
\bibfield{author}{\bibinfo{person}{Camille~L Ryan} {and} \bibinfo{person}{Kurt
  Bauman}.} \bibinfo{year}{2016}\natexlab{}.
\newblock \bibinfo{booktitle}{\emph{Educational attainment in the United
  States: 2015}}.
\newblock \bibinfo{type}{{T}echnical {R}eport}. \bibinfo{institution}{U.S.
  Census Bureau}.
\newblock


\bibitem[\protect\citeauthoryear{Schaefer, Simpkins, Vest, and Price}{Schaefer
  et~al\mbox{.}}{2011}]%
        {schaefer11}
\bibfield{author}{\bibinfo{person}{David~R Schaefer}, \bibinfo{person}{Sandra~D
  Simpkins}, \bibinfo{person}{Andrea~E Vest}, {and} \bibinfo{person}{Chara~D
  Price}.} \bibinfo{year}{2011}\natexlab{}.
\newblock \showarticletitle{The Contribution of Extracurricular Activities to
  Adolescent Friendships: New Insights through Social Network Analysis}.
\newblock \bibinfo{journal}{\emph{Developmental psychology}}
  \bibinfo{volume}{47}, \bibinfo{number}{4} (\bibinfo{year}{2011}),
  \bibinfo{pages}{1141--1152}.
\newblock


\bibitem[\protect\citeauthoryear{Shalizi and Rinaldo}{Shalizi and
  Rinaldo}{2013}]%
        {shalizi13}
\bibfield{author}{\bibinfo{person}{Cosma~Rohilla Shalizi} {and}
  \bibinfo{person}{Alessandro Rinaldo}.} \bibinfo{year}{2013}\natexlab{}.
\newblock \showarticletitle{{Consistency under sampling of exponential random
  graph models}}.
\newblock \bibinfo{journal}{\emph{The Annals of Statistics}}
  \bibinfo{volume}{41}, \bibinfo{number}{2} (\bibinfo{year}{2013}),
  \bibinfo{pages}{508--535}.
\newblock


\bibitem[\protect\citeauthoryear{Traud, Kelsic, Mucha, and Porter}{Traud
  et~al\mbox{.}}{2011}]%
        {traud11}
\bibfield{author}{\bibinfo{person}{Amanda~L Traud}, \bibinfo{person}{Eric~D
  Kelsic}, \bibinfo{person}{Peter~J Mucha}, {and} \bibinfo{person}{Mason~A
  Porter}.} \bibinfo{year}{2011}\natexlab{}.
\newblock \showarticletitle{{Comparing community structure to characteristics
  in online collegiate social networks}}.
\newblock \bibinfo{journal}{\emph{SIAM Rev.}} \bibinfo{volume}{53},
  \bibinfo{number}{3} (\bibinfo{year}{2011}), \bibinfo{pages}{526--543}.
\newblock


\bibitem[\protect\citeauthoryear{Traud, Mucha, and Porter}{Traud
  et~al\mbox{.}}{2012}]%
        {traud12}
\bibfield{author}{\bibinfo{person}{Amanda~L Traud}, \bibinfo{person}{Peter~J
  Mucha}, {and} \bibinfo{person}{Mason~A Porter}.}
  \bibinfo{year}{2012}\natexlab{}.
\newblock \showarticletitle{{Social structure of Facebook networks}}.
\newblock \bibinfo{journal}{\emph{Physica A}} \bibinfo{volume}{391},
  \bibinfo{number}{16} (\bibinfo{date}{Aug.} \bibinfo{year}{2012}),
  \bibinfo{pages}{4165--4180}.
\newblock


\bibitem[\protect\citeauthoryear{Ugander, Backstrom, Marlow, and
  Kleinberg}{Ugander et~al\mbox{.}}{2012}]%
        {ugander12}
\bibfield{author}{\bibinfo{person}{Johan Ugander}, \bibinfo{person}{Lars
  Backstrom}, \bibinfo{person}{Cameron Marlow}, {and} \bibinfo{person}{Jon
  Kleinberg}.} \bibinfo{year}{2012}\natexlab{}.
\newblock \showarticletitle{Structural diversity in social contagion}.
\newblock \bibinfo{journal}{\emph{Proceedings of the National Academy of
  Sciences}} \bibinfo{volume}{109}, \bibinfo{number}{16}
  (\bibinfo{year}{2012}), \bibinfo{pages}{5962--5966}.
\newblock


\bibitem[\protect\citeauthoryear{Van~Duijn, Zeggelink, Huisman, Stokman, and
  Wasseur}{Van~Duijn et~al\mbox{.}}{2003}]%
        {van03}
\bibfield{author}{\bibinfo{person}{Marijtje Van~Duijn},
  \bibinfo{person}{Evelien Zeggelink}, \bibinfo{person}{Mark Huisman},
  \bibinfo{person}{Frans Stokman}, {and} \bibinfo{person}{Frans Wasseur}.}
  \bibinfo{year}{2003}\natexlab{}.
\newblock \showarticletitle{Evolution of sociology freshmen into a friendship
  network}.
\newblock \bibinfo{journal}{\emph{Journal of Mathematical Sociology}}
  \bibinfo{volume}{27}, \bibinfo{number}{2-3} (\bibinfo{year}{2003}),
  \bibinfo{pages}{153--191}.
\newblock


\bibitem[\protect\citeauthoryear{Viswanath, Mislove, Cha, and
  Gummadi}{Viswanath et~al\mbox{.}}{2009}]%
        {viswanath09}
\bibfield{author}{\bibinfo{person}{Bimal Viswanath}, \bibinfo{person}{Alan
  Mislove}, \bibinfo{person}{Meeyoung Cha}, {and} \bibinfo{person}{Krishna~P
  Gummadi}.} \bibinfo{year}{2009}\natexlab{}.
\newblock \showarticletitle{On the evolution of user interaction in facebook}.
  In \bibinfo{booktitle}{\emph{Proceedings of the 2nd ACM workshop on Online
  social networks}}. \bibinfo{pages}{37--42}.
\newblock


\bibitem[\protect\citeauthoryear{Watts and Strogatz}{Watts and
  Strogatz}{1998}]%
        {watts98}
\bibfield{author}{\bibinfo{person}{Duncan~J Watts} {and}
  \bibinfo{person}{Steven~H Strogatz}.} \bibinfo{year}{1998}\natexlab{}.
\newblock \showarticletitle{Collective dynamics of ‘small-world’networks}.
\newblock \bibinfo{journal}{\emph{Nature}} \bibinfo{volume}{393},
  \bibinfo{number}{6684} (\bibinfo{year}{1998}), \bibinfo{pages}{440}.
\newblock


\bibitem[\protect\citeauthoryear{Wimmer and Lewis}{Wimmer and Lewis}{2010}]%
        {wimmer10}
\bibfield{author}{\bibinfo{person}{Andreas Wimmer} {and} \bibinfo{person}{Kevin
  Lewis}.} \bibinfo{year}{2010}\natexlab{}.
\newblock \showarticletitle{{Beyond and below racial homophily: ERG models of a
  friendship network documented on Facebook.}}
\newblock \bibinfo{journal}{\emph{Amer. J. Sociology}} \bibinfo{volume}{116},
  \bibinfo{number}{2} (\bibinfo{date}{Sept.} \bibinfo{year}{2010}),
  \bibinfo{pages}{583--642}.
\newblock


\end{thebibliography}

\end{document}